\documentclass[onecolumn]{pasj00}
\draft

\newcommand{\marka}{\footnotemark[*]}
\newcommand{\markb}{\footnotemark[$\dagger$]}
\newcommand{\markc}{\footnotemark[$\ddagger$]}
\newcommand{\markd}{\footnotemark[$\S$]}
\newcommand{\marke}{\footnotemark[$\|$]}
\newcommand{\markf}{\footnotemark[$\#$]}
\newcommand{\markg}{\footnotemark[**]}
\newcommand{\markh}{\footnotemark[$\dagger \dagger$]}
\newcommand{\marki}{\footnotemark[$\ddagger \ddagger$]}
\newcommand{\markj}{\footnotemark[$\S \S$]}
\newcommand{\markk}{\footnotemark[$\| \|$]}
\newcommand{\markl}{\footnotemark[$\# \#$]}
\newcommand{\markm}{\footnotemark[***]}
\newcommand{\markn}{\footnotemark[$\dagger \dagger \dagger$]}
\newcommand{\marko}{\footnotemark[$\ddagger \ddagger \ddagger$]}
\newcommand{\markp}{\footnotemark[$\S \S \S$]}
\newcommand{\markq}{\footnotemark[$\| \| \|$]}
\newcommand{\markr}{\footnotemark[$\# \# \#$]}

\begin{document}
\SetRunningHead{K. Imanishi et al.}{X-Ray Flares from Low-Mass YSOs in the $\rho$ Oph Star-Forming Region}

\Received{2002/11/23}
\Accepted{2003/03/12}

\title{A Systematic Study of X-Ray Flares from Low-Mass Young Stellar
Objects in the $\rho$ Ophiuchi Star-Forming Region with Chandra}

\author{Kensuke \textsc{Imanishi}, Hiroshi \textsc{Nakajima}, Masahiro
\textsc{Tsujimoto}, \\ Katsuji \textsc{Koyama}}%
\affil{Department of Physics, Graduate School of Science, Kyoto
University, \\ Sakyo-ku, Kyoto. 606-8502}
\email{kensuke@cr.scphys.kyoto-u.ac.jp,
nakajima@cr.scphys.kyoto-u.ac.jp, \\ tsujimot@cr.scphys.kyoto-u.ac.jp,
koyama@cr.scphys.kyoto-u.ac.jp}
\and
\author{Yohko \textsc{Tsuboi}}
\affil{Department of Science and Engineering, Chuo University,
Bunkyo-ku, Tokyo. 112-8551}
\email{tsuboi@phys.chuo-u.ac.jp}

%


\maketitle

\begin{abstract}
 We report on the results of a systematic study of X-ray flares from
 low-mass young stellar objects, using two deep exposure Chandra
 observations of the main region of the $\rho$ Ophiuchi star-forming
 cloud. From 195 X-ray sources, including class I--III sources and some
 young brown dwarfs, we detected a total of 71 X-ray flares. Most of the
 flares have the typical profile of solar and stellar flares, fast rise
 and slow decay, while some bright flares show unusually long rise
 timescales. We derived the time-averaged temperature ($\langle kT\rangle$),
 luminosity ($\langle L_{\rm X}\rangle$), rise and decay timescales
 ($\tau_{\rm r}$ and $\tau_{\rm d}$) of the flares, finding that
 (1) class I--II sources tend to have a high $\langle kT\rangle$,
 which sometimes exceeds 5 keV, 
%
%
 (2) the distribution of $\langle L_{\rm X}\rangle$ during flares is nearly
 the same for all classes from $\sim$10$^{29.5}$ to $\sim$10$^{31.5}$ erg
 s$^{-1}$, although there is a marginal hint of a higher
 $\langle L_{\rm X}\rangle$ distribution for class I than class II--III, and
 (3) positive and negative log-linear correlations are found between
 $\tau_{\rm r}$ and $\tau_{\rm d}$, and $\langle kT\rangle$ and $\tau_{\rm r}$.
 In order to explain these relations, we used the framework of magnetic
 reconnection model with heat conduction and chromospheric evaporation
 to formulate the observational
 parameters ($\tau_{\rm r}$, $\tau_{\rm d}$, and $\langle kT\rangle$) as
 a function of the pre-flare (coronal) electronic density ($n_{\rm c}$),
 the half-length of the reconnected magnetic loop ($L$),
 and magnetic field strength ($B$). The observed correlations are
 well reproduced if loop lengths are nearly
 the same for all classes, regardless of the existence of an accretion
 disk. The estimated loop length is almost comparable to the typical
 stellar radius of these objects (10$^{10}$--10$^{11}$ cm), which
 indicates that the observed flares are triggered by solar-type loops,
 rather than larger ones ($\sim$10$^{12}$ cm) connecting the star with
 its inner accretion disk. The higher $\langle kT\rangle$ observed for class I
 sources may be explained by a slightly higher magnetic field strength
 ($\approx$500 G) than for class II--III sources (200--300 G).
\end{abstract}

\section{Introduction}
\label{sec:intro}

Low-mass young stellar objects (YSOs) are classified into four
evolutional stages based on the infrared (IR) to sub-millimeter spectral
energy distributions (SEDs); the youngest and evolved protostars have
class 0 and I SEDs, while classical and weak-lined T Tauri stars (CTTSs
and WTTSs) exhibit class II and III SEDs, respectively \citep{Lada1991,
Andre1993, Andre1994}.
In the 1980's, the Einstein satellite discovered X-rays from T Tauri
stars (TTSs = CTTSs and WTTSs: \cite{Feigelson1981, Feigelson1981b,
Montmerle1983}).
%
%
\citet{Koyama1994} first detected X-rays from the positions of class I
sources in the $\rho$ Ophiuchi dark cloud (WL 6 and WL 15 = Elias 29) with
ASCA/GIS.  Using ASCA/SIS, which has a better spatial
resolution than GIS, \citet{Kamata1997} confirmed the X-ray emission
from the class I sources.  Independently, \citet{Casanova1995} obtained
hints of X-ray detection from class I in the same star-forming area
with ROSAT/PSPC. Then, successive observations with ASCA
further revealed that many class I protostars are X-ray emitters
\citep{Koyama1996, Ozawa1999, Tsuboi2000}, thanks to a higher
sensitivity for hard X-ray photons ($>$2 keV), which are less
absorbed, because the extinction cross-section decreases as
$E_\mathrm{X}^{-2.5}$ ($E_\mathrm{X}$: X-ray energy). ROSAT
observations also detected X-rays from class I \citep{Grosso1997,
Neuhauser1997, Grosso2001}.
They share the same characteristics of thermal emission with a plasma
temperature of 0.5--5 keV and a strong variability with occasional rapid
flares, consistent with the scenario of enhanced solar-type activity,
attributable to magnetic dynamo processes.
%
%
Furthermore, a recent observation with the Chandra satellite
detected hard X-rays from class 0 source candidates in the Orion
Molecular Cloud 3 \citep{Tsuboi2001}, although its characteristics are
still poorly understood due to the limited statistics.
These pioneering discoveries demonstrate the unique capability of hard
X-ray observations to probe YSOs or their close vicinity deeply embedded
in dense cores.

Based on the above results, the next step should be a systematic study
of X-rays from YSOs and to approach physical conditions of these initial
stages of stars, which gives extremely important information concerning
star-formation theory.
For example, our earlier results of a Chandra observation of the
$\rho$ Ophiuchi cloud suggested that class I protostars tend to show a
higher plasma temperature ($kT$) than TTSs (Imanishi et~al. 2001a,
hereafter Paper I). 
%
%
\citet{Stelzer2000} also did a systematic study of TTS flares in
Taurus--Auriga--Perseus with ROSAT/PSPC, and found that the flare
rate of CTTSs may be somewhat higher than that of WTTSs.
%
%
To explain the quasi-periodic X-ray flare observed from the class I
source YLW 15 in the $\rho$ Ophiuchi dark cloud \citep{Tsuboi2000},
\citet{Montmerle2000} proposed a scenario of star--disk connection
by one magnetic loop, where the central forming-star rotates faster than
the inner edge of its accretion disk, which triggers periodic X-ray
flaring.
Shibata and Yokoyama (1999, 2002) further showed that $kT$ is determined
by the balance between reconnection heating and conduction cooling, and then
derived the relation $kT \propto L^{2/7} B^{6/7}$, where $L$ and $B$
are the half-length of the reconnected magnetic loop and strength of
the magnetic field. Hence, larger $L$ and/or $B$ values make a higher
temperature plasma.
Combining these results, one plausible scenario is that younger sources
have a much larger flare loop than evolved sources connecting the central
source and the disk, and thus show frequent flares with a higher plasma
temperature.
\citet{Feigelson2002}, on the other hand, reported that X-ray emission
has no correlation with the presence or absence of the disk for the
Orion Nebula Cluster X-ray sources. It is therefore still controversial
whether there is any differences in the observed X-ray properties or
emission mechanisms between the different classes.

To address this issue, other important parameters would be the
timescales of the flares, which were not considered in the above
results. If we assume radiative cooling without successive heating
\citep{vandenOord1988}, the decay timescale of flares ($\tau_{\rm d}$) is
equal to the radiative loss timescale, defined as $E_{\rm{th}}$/$R$
($E_{\rm{th}}$
is the total thermal energy and $R$ is emissivity of the plasma). Since
$E_{\rm{th}}$ is supplied by the magnetic energy ($B^2/8\pi$) and $R$ depends
on their plasma electronic density ($R \propto n^2$, $n$: electronic
density), $\tau_{\rm d}$ is tightly correlated with $B$ and $n$.
Also, a standard magnetic reconnection model \citep{Petschek1964}
predicted that the rise timescale of flares ($\tau_{\rm r}$) is proportional
to the Alfv\'en time ($\tau_{\rm A}$), defined as $L$/$v_{\rm A}$, where
$v_{\rm A}$
is the Alfv\'en velocity ($v_{\rm A} \propto B$); hence, a larger $L$ and/or
a smaller $B$ yield a longer $\tau_{\rm r}$.
In fact, some stellar flares display unusually long rising phases (e.g.,
UX Ari: \cite{Gudel1999} and ROXs31: \cite{Imanishi2002}).
To examine any correlations between these observable parameters ($kT$,
$\tau_{\rm r}$, and $\tau_{\rm d}$) would be fruitful for understanding of physical
conditions of YSOs. However, little has been done for systematical approach
so far.

In the present work, we made the first systematic study of the 
flare activity of
YSOs, using two deep exposure Chandra observations of the $\rho$
Ophiuchi cloud (hereafter, $\rho$ Oph) at a distance of 145 pc
\citep{deZeeuw1999}. Details of the observations are shown in section
\ref{sec:obs}. We then performed timing and spectral analyses for all
detected sources, and examine any correlations between the derived parameters
(section \ref{sec:analysis_results}). From the observed quantities, we estimate
the physical parameters and discuss the differences of the derived
parameters between each class (section \ref{sec:discussion}).

\section{Chandra Observations of the $\rho$ Ophiuchi Cloud}
\label{sec:obs}

The Chandra X-ray Observatory \citep{Weisskopf2002} observed the
central region of $\rho$ Oph twice with a deep exposure of the ACIS-I
array, consisting of four abutted X-ray CCDs. The first observation (here
and after, obs.-BF) covered the south-east 17\farcm4$\times$17\farcm4
area, including cores B, C, E, and F, while the second observation
(obs.-A) covered the north-west area centered on core A
\citep{Loren1990}. Although some of the ACIS-S chips were simultaneously
in operation, we do not use these data because large off-axis angles
cause a degeneration of the sensitivity and position determination. From the
Chandra X-ray Center (CXC) archive, we retrieved level-2 data,
in which the data degradation caused by the increase of charge transfer
inefficiency (CTI) in orbit was corrected. The X-ray events were
selected with ASCA grades 0, 2, 3, 4, and 6. The afterglow
events were also removed. After processing, each observation yielded
$\approx$100 ks of live time (table \ref{tab:obs}). Earlier results of
obs.-BF were found in Paper I.

\section{Analyses and Results}
\label{sec:analysis_results}

\subsection{X-ray Sources and NIR Counterparts}
\label{ssec:src}

\begin{figure}[htbp]
 \begin{center}
  \FigureFile(180mm,180mm){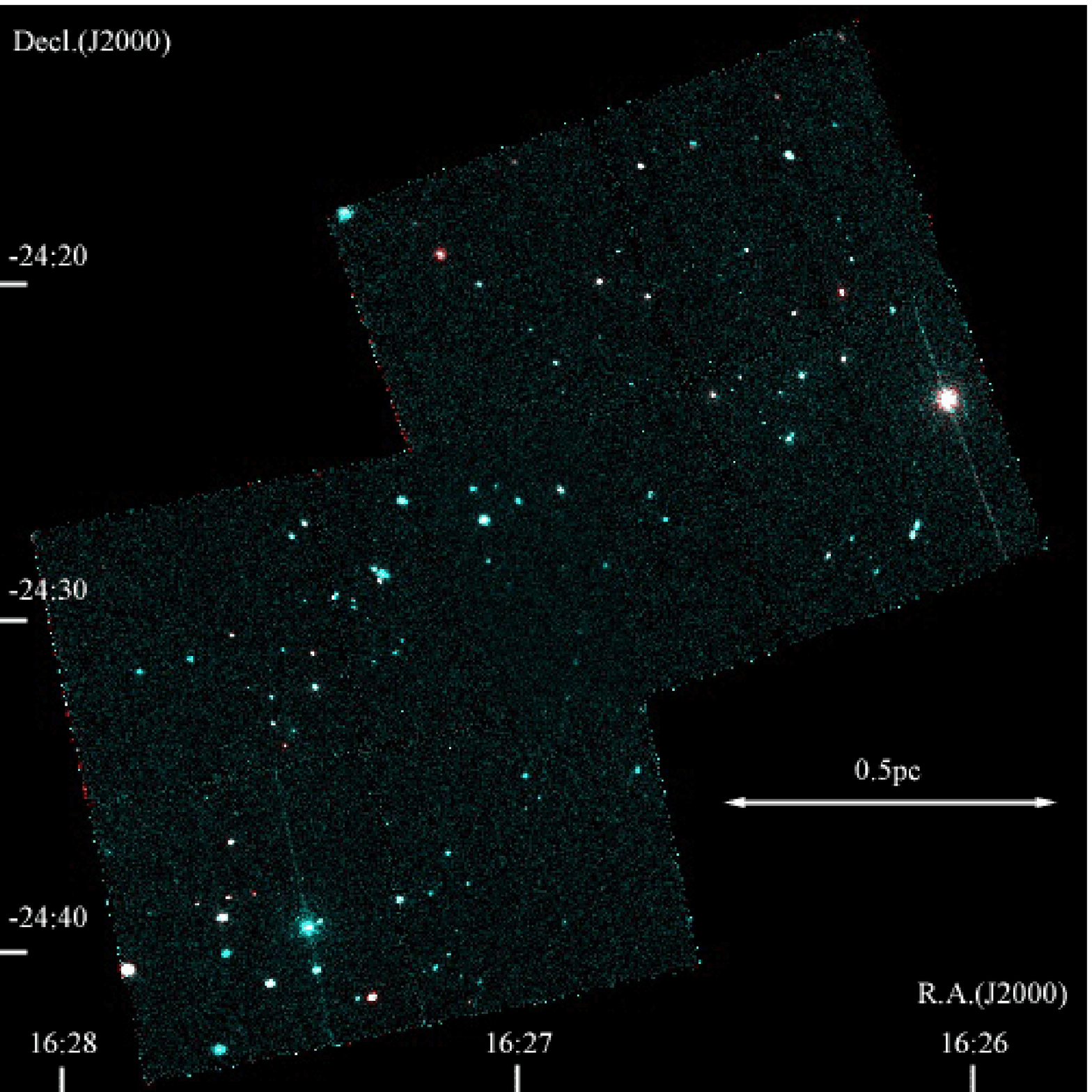}
 \end{center}
 \caption{False-color ACIS image of $\rho$ Oph. Red and blue colors
 represent photons in the soft (0.5--2.0~keV) and hard (2.0--9.0~keV)
 bands, respectively. \label{fig:img}}
\end{figure}

Figure \ref{fig:img} gives an ACIS image of $\rho$ Oph. The red and blue
colors represent photons in the soft (0.5--2.0~keV) and hard
(2.0--9.0~keV) X-ray bands, respectively.
First, we discuss the source detection analysis using the wavdetect command
\citep{Freeman2002} in the CIAO package for 2048$\times$2048 pixel
images with the pixel size of $\sim$0\farcs5. We then consider the search for
near-infrared (NIR) counterparts for all detected X-ray sources.
The basic procedures of the analysis were the same in Paper I, except for the
following points:\\
(1)~The significance criterion of wavdetect is relaxed from 10$^{-7}$
(Paper I) to 10$^{-6}$.\\
(2)~We apply the wavdetect analysis for both the soft and hard X-ray
bands as well as for the total (0.5--9.0~keV) X-ray band.\\
(3)~For reference of NIR sources, we use the Point Source Catalog in
the 2MASS Second Incremental Data Release\footnote{See $\langle$
http://www.ipac.caltech.edu/2mass/releases/second/doc/$\rangle$.}, not that in
\citet{Barsony1997}, because the former has a better position accuracy
($\sigma \sim$0\farcs1) than, and comparable sensitivity ($Ks <$ 14.3
mag) to the latter (\cite{Barsony1997}, $\sigma \sim$1\farcs2 and $K <$
14.5 mag).

We then detected 195 X-ray sources, nine of which were found in both the
two observations. Thirteen and nine sources were exclusively found in
only the soft and hard bands, respectively (hereafter, the ``soft-band
sources'' and ``hard-band sources''). Table \ref{tab:src} (columns
1--4) gives the name, background-subtracted ACIS-I counts, and the
coordinates (after the offset correction) for each source. The soft-band
and hard-band sources are indicated by the prefixes ``S'' and ``H''.  
The X-ray photons were extracted from a circle of 1\farcs2--23\farcs3
radius, depending on the point spread function (PSF) radius for 1.49 keV
photons, which is a function of the angular distance from the optical
axis of the telescope 
%
%
(psfsize20010416.fits\footnote{This file was constructed by using the
SAOsac raytrace code (\cite{Jerius1995}, see also
http://asc.harvard.edu/chart/).} in the CIAO 2.2.1 package). Typically,
in the source radius, $\approx$95\% of the X-ray photons are included. In order
to obtain a high signal-to-noise ratio for some faint sources, we used a
circle of a half radius of the PSF. Background counts were extracted from
circles of 38 arcmin$^2$ and 57 arcmin$^2$ areas from the respective
source-free regions in the ACIS-I fields of obs.-A and obs.-BF.

In order to check for possible false sources with the wavdetect procedure
\citep{Feigelson2002}, we estimated the confidence level ($CL$) of the
X-ray counts using Poisson statistics \citep{Imanishi2001b}. Then, 14
sources were found to have a $CL$ value smaller than 99.9\%.
Although A-48 and A-H2 have
significantly high $CL$, these may also be possible false sources
because of the larger source size than the PSF radius (A-48) and of the
severe contamination from A-2 (A-H2). We hence note that 16 sources were
marginal detections, and label them with (m) in column 2 of table
\ref{tab:src}.

About 60\% of the X-ray sources had 2MASS NIR counterparts. Using these
pairs, we determined and shifted the absolute positions of the
Chandra sources so that the mean values of the Chandra-2MASS
offset in the direction of right ascension and declination would become
zero. The offset correction of obs.-BF (hence the position of the X-ray
sources) was slightly different from those reported in Paper I because of
the difference of the reference NIR catalogue.
For the remaining sources, we further searched for counterparts using
other NIR catalogues, and then identified an X-ray source, BF-90, to be a
counterpart of GY 322 \citep{Greene1992}. Finally, we concluded that 110
out of 195 X-ray sources have NIR counterparts. The offset between the
X-ray and NIR sources is shown at column 2 in table \ref{tab:id}. We
also searched for radio (cm) and X-ray counterparts in published
catalogues (columns 3--7 in table \ref{tab:id}).

Column 9 in table \ref{tab:id} gives the IR classification based on
the spectral indices from the NIR to mid-IR band (ISOCAM survey at 6.7
and 14.3 $\mu$m; \cite{Bontemps2001}). \citet{Imanishi2001a} used the
terminology class I$_{\rm c}$ (class I candidate) for sources previously
classified as class I (or flat spectrum source). In this paper, however,
we regard all class I$_{\rm c}$s as class II, following \citet{Bontemps2001}.
Furthermore, we define some additional classes (class III$_{\rm c}$, BD,
BD$_{\rm c}$, F, unclassified NIR sources, and unidentified sources). The
definition of these classes is given in \citet{Bontemps2001},
\citet{Imanishi2001b}, and a footnote of table \ref{tab:id} of this
paper.
We then list 8 class Is, 58 class IIs, 17 class IIIs, 9 class III$_{\rm c}$s
(class III candidates), 2 BDs (brown dwarfs), 3 BD$_{\rm c}$s (BD candidates),
1 F (foreground star), and 12 unclassified NIR sources.
Hereafter, we use the terminology ``class III+III$_{\rm c}$'' by combining the
class III and class III$_{\rm c}$ for brevity.

\subsection{Timing Analysis}
\label{sec:timing}

\begin{figure}[htbp]
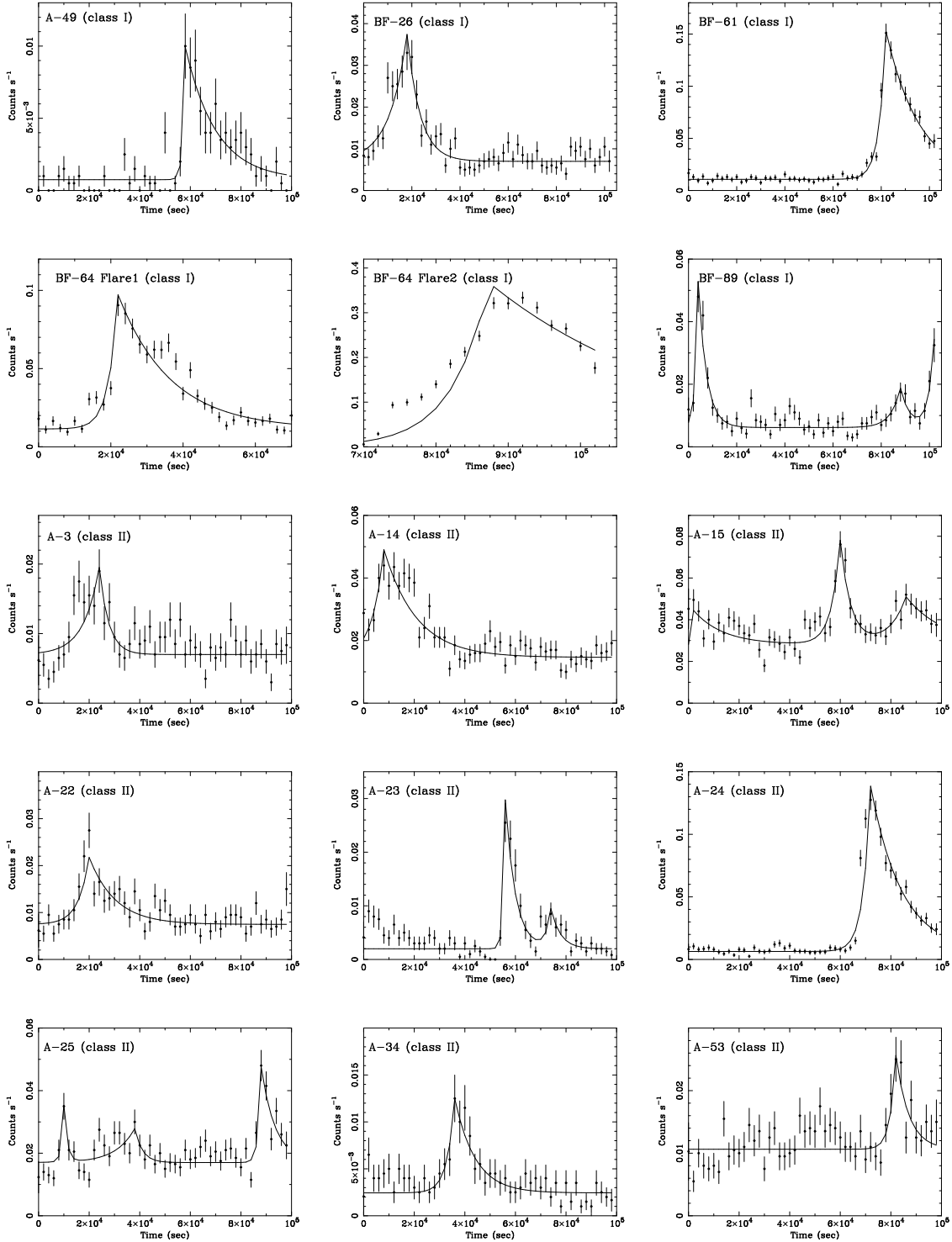

 \begin{center}
  \FigureFile(50mm,40mm){figure2_01.eps}
  \FigureFile(50mm,40mm){figure2_02.eps}
  \FigureFile(50mm,40mm){figure2_03.eps}
  \FigureFile(50mm,40mm){figure2_04.eps}
  \FigureFile(50mm,40mm){figure2_05.eps}
  \FigureFile(50mm,40mm){figure2_06.eps}
  \FigureFile(50mm,40mm){figure2_07.eps}
  \FigureFile(50mm,40mm){figure2_08.eps}
  \FigureFile(50mm,40mm){figure2_09.eps}
  \FigureFile(50mm,40mm){figure2_10.eps}
  \FigureFile(50mm,40mm){figure2_11.eps}
  \FigureFile(50mm,40mm){figure2_12.eps}
  \FigureFile(50mm,40mm){figure2_13.eps}
  \FigureFile(50mm,40mm){figure2_14.eps}
  \FigureFile(50mm,40mm){figure2_15.eps}
 \end{center}
 \caption{Light curves of all detected flares in 0.5--9.0 keV, binned in
 2000 s interval. Names and classes for each flare are shown at the
 upper-left of the panels. The time axis starts at MJD = 51679.9944
 (obs.-A) and 51647.7848 (obs.-BF). The solid line is the best-fit
 (exponential + constant) model. Since the light curve of BF-64 suffers
 photon pileup during the second flare, we show the two flares
 separately (the former is extracted from the 7\farcs5 radius and the
 latter is from 2\farcs5--7\farcs5). That of A-2 is also extracted from
 the 2\farcs5--12\farcs5 radius circle. \label{fig:flares}}
\end{figure}

We made two X-ray light curves in the 0.5--9.0 keV region for all of the
Chandra sources (background was not subtracted), with respective time bins of
2000 s and 4000 s. The source regions were basically the same as those used
in an estimation of the source counts (subsection \ref{ssec:src}). For the
brightest two sources (A-2 and BF-64), however, we used an annulus of
2\farcs5--12\farcs5 and 2\farcs5--7\farcs5 radius in order to avoid
photon pileup \citep{Davis2001}. The light curves show many flare-like
events. We defined a ``flare time bin'' by the following criterion:
\begin{equation}
 \label{eq:flare}
 \frac{N - N_0}{\Delta N} \geq 2,
\end{equation}
where $N$, $\Delta N$, and $N_0$ are the X-ray counts, the statistical
1-$\sigma$ error for the relevant time bin, and average counts in a
20000 s interval including the relevant time bin, respectively. If the
above criterion was satisfied in both of the two light curves (2000 s and
4000 s time bin), we defined the event to be a ``flare''. We then picked
up 71 flares (table \ref{tab:flare}). Figure \ref{fig:flares} shows all of the
light curves of the detected flares. We note that the flare list in
Paper I is slightly different from that in this paper, due solely to the
different definition of a flare. This slight difference, however, has
no significant effect on the following analyses and discussions.

Most flares had the typical profile of those from low-mass
main-sequence stars or YSOs; fast-rise and slow-decay, while some
sources show unusual flares having slow rise timescale; e.g., BF-64 =
YLW 16A (figure \ref{fig:flares}).
We fit the rise and decay phases of the flare by a simple (exponential +
constant) model using the QDP command in the LHEASOFT 5.0\footnote{See
$\langle$ http://heasarc.gsfc.nasa.gov/docs/software/lheasoft/$\rangle$.},
and derived the
respective $e$-folding times and quiescent level ($\tau_{\rm r}$, $\tau_{\rm d}$,
and $Q$; table \ref{tab:flare}) with their 90\% errors, where the flare
peak time ($t_{\rm p}$) was fixed to be the maximum time bin, except for the
second flare of BF-64.
Evidently, this simple model can not reproduce the unusually giant flare
from BF-64, even if we relax $t_{\rm p}$ to be free. However, we do not intend
to make a model of the flare light curve and only have interest in the
typical timescales in this paper; hence, we do not discuss this discrepancy
in further detail.

\subsection{Spectral Analysis}
\label{sec:spec}

X-ray spectra were made for all of the Chandra sources. For
flaring sources, we made time-averaged spectra in the quiescent and
flare phases separately. The flare phase was the time from ($t_{\rm p}$ $-$
$\tau_{\rm r}$) to ($t_{\rm p}$ + $\tau_{\rm d}$), and the quiescent phase was
that outside of the flare phase. The background regions were the same as
used in
subsection \ref{ssec:src}. We then fit the spectra by a thin-thermal plasma
model (MEKAL: \cite{Mewe1985}) with the photoelectric absorption
(WABS). The metal abundances were fixed to be 0.3 solar based on
previous fitting results (Paper I), unless otherwise noted. If the
temperature was not constrained, which is often the case for very faint
sources, we fixed the temperature at two representative values of 1 keV
(typical of TTSs) and 5 keV (protostars), then estimated the respective
absorptions and luminosities. For the nine sources detected in both of the
observations, we derived the parameters of each spectrum assuming the
same absorption. This simple model was generally acceptable. BF-46
(ROXs21) and BF-96 (ROXs31), however, needed multi-temperature spectra
with unusual abundances \citep{Imanishi2002}. A-23, A-24 and BF-10, on
the other hand, required overabundances (subsubsection 4.8.1 in Paper I;
\cite{Imanishi_phD}). In table \ref{tab:src} (columns 5--9), we
summarize the best-fit parameters; time-averaged temperature
($\langle kT\rangle$), emission measure ($\langle EM\rangle$),
absorption column ($N_{\rm H}$), flux, and
luminosity ($\langle L_{\rm X}\rangle$) in 0.5--9.0~keV.

\subsection{Luminosity Function}
\label{ssec:lx}

\begin{figure}[htbp]
 \begin{center}
  \FigureFile(80mm,50mm){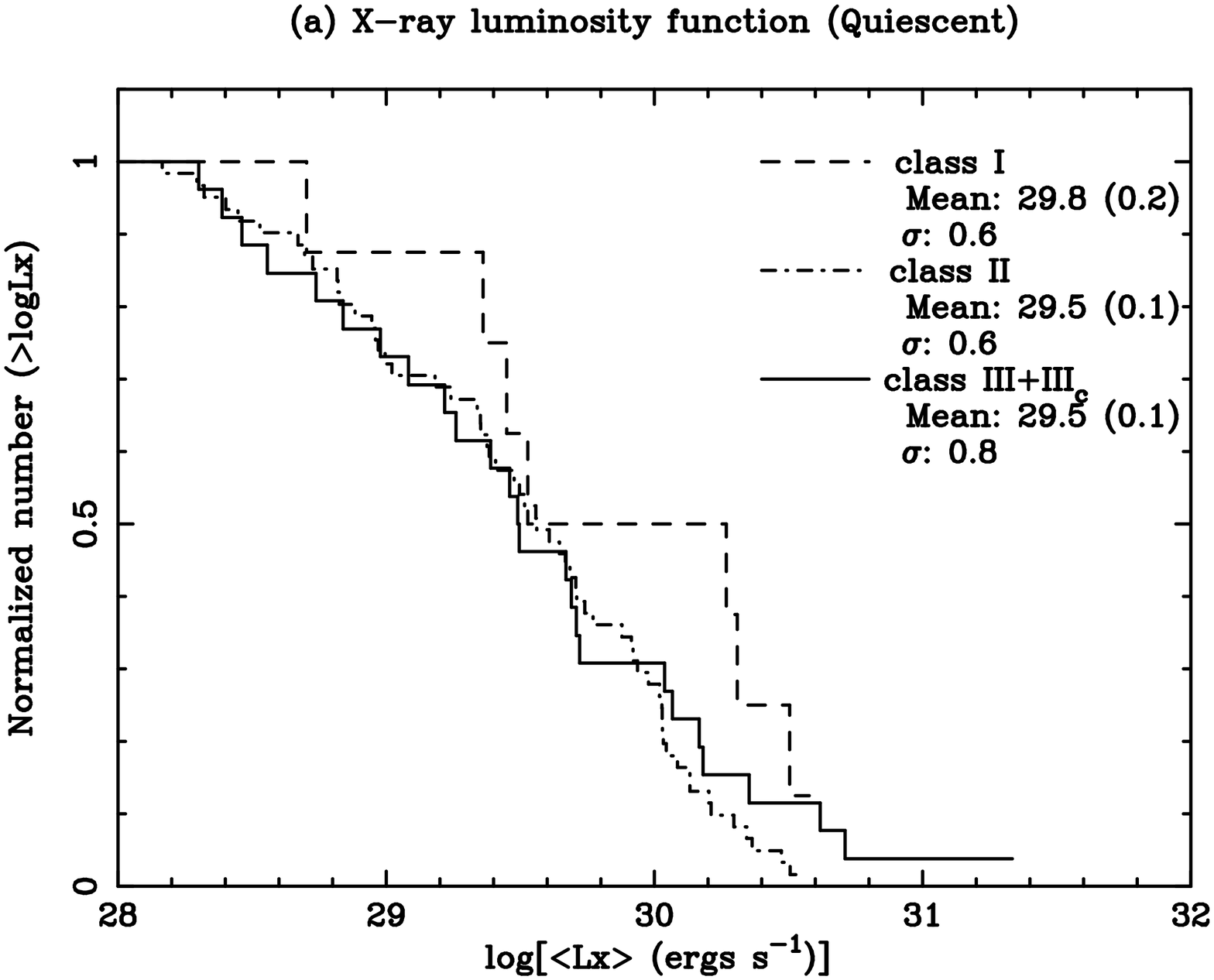}
  \FigureFile(80mm,50mm){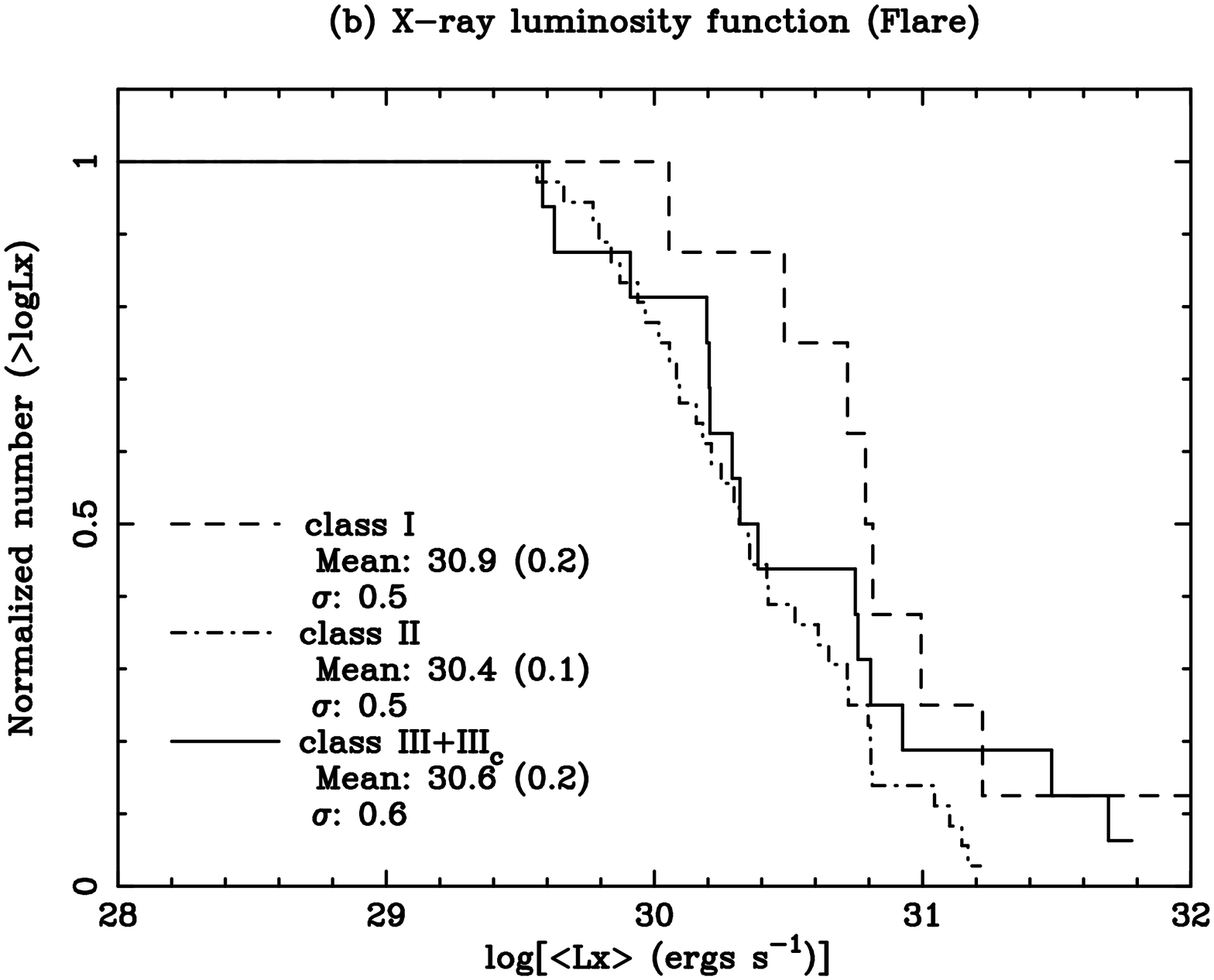}
 \end{center}
 \caption{Normalized X-ray luminosity functions of class I (dashed),
 class II (dash-dotted), and class III (solid) sources in the (a)
 quiescent and (b) flare phases. The mean value (Mean) and standard
 deviation ($\sigma$) of log[$L_{\rm X}$] in the unit of erg s$^{-1}$ for
 each class are shown in the figures. The parentheses indicate errors of
 the mean values. \label{fig:lx_func}}
\end{figure}

In order to examine the differences along the evolutional stages, we
calculated the X-ray luminosity function for the detected X-ray sources
of class I, II, and III+III$_{\rm c}$ (figure \ref{fig:lx_func}), using the
ASURV statistical software package (rev.1.2)\footnote{See
$\langle$http://www.astro.psu.edu/statcodes$\rangle$.}
based on the maximum likelihood Kaplan-Meier estimator.
The luminosity function of class I seems to be shifted toward higher
values than those of class II and III+III$_{\rm c}$ both in the quiescent and
flare phases.

%
%
To be more quantitative, we estimated the significance level using
two nonparametric two-sample tests in ASURV: the Gehan's generalized
Wilcoxon test (GW) and the logrank test \citep{Feigelson1985}. The
results are given in table \ref{tab:asurv}, where the blank (...)
indicates that the significance level is less than 90\%.
From table \ref{tab:asurv}, we can see that both the GW and logrank
tests for the $\langle L_{\rm X}\rangle$ difference between class I and
the others in the flares show a marginal significance level of $\sim$94\%.
We should further note that the effect of ``undetected'' faint flares of
class I can not be ignored; if we assume an undetected class I has
$\langle L_{\rm X}\rangle$ smaller than 10$^{30}$ erg~s$^{-1}$, which is
near to the detection
threshold of a class I flare, and add this source to the GW and logrank
test sample, then the relevant significance level is largely reduced to
$<$ 90\%.  We hence conclude that the higher $\langle L_{\rm X}\rangle$ of
class I than class II+III during flares is marginal.

On the other hand, from table \ref{tab:asurv}, we can see no significant
difference between class II and III+III$_{\rm c}$ source in the flare
phase. This is contrary to the results of the Taurus--Auriga--Perseus
samples \citep{Stelzer2000}, for which there was a significant
difference between the distribution of flare $\langle L_{\rm X}\rangle$ of
class II (CTTS) and class III+III$_{\rm c}$ (WTTS) sources.

Also, we can see no significant difference of $\langle L_{\rm X}\rangle$
in the quiescent phase
among all classes. This is consistent with the previous estimation
for this region with ROSAT derived by \citet{Grosso2000}. They
tried to estimate the significances rather strictly by considering the
upper limit of undetected sources. Our samples give a more severe
constraint because the detection threshold is largely reduced to give
mean luminosities of 10$^{29.5}$--10$^{29.8}$ erg s$^{-1}$, which
are about 10-times lower than that of ROSAT.

\subsection{Plasma Temperature and Flare Timescales}
\label{ssec:para}

\begin{figure}[htbp]
 \begin{center}
  \FigureFile(50mm,80mm){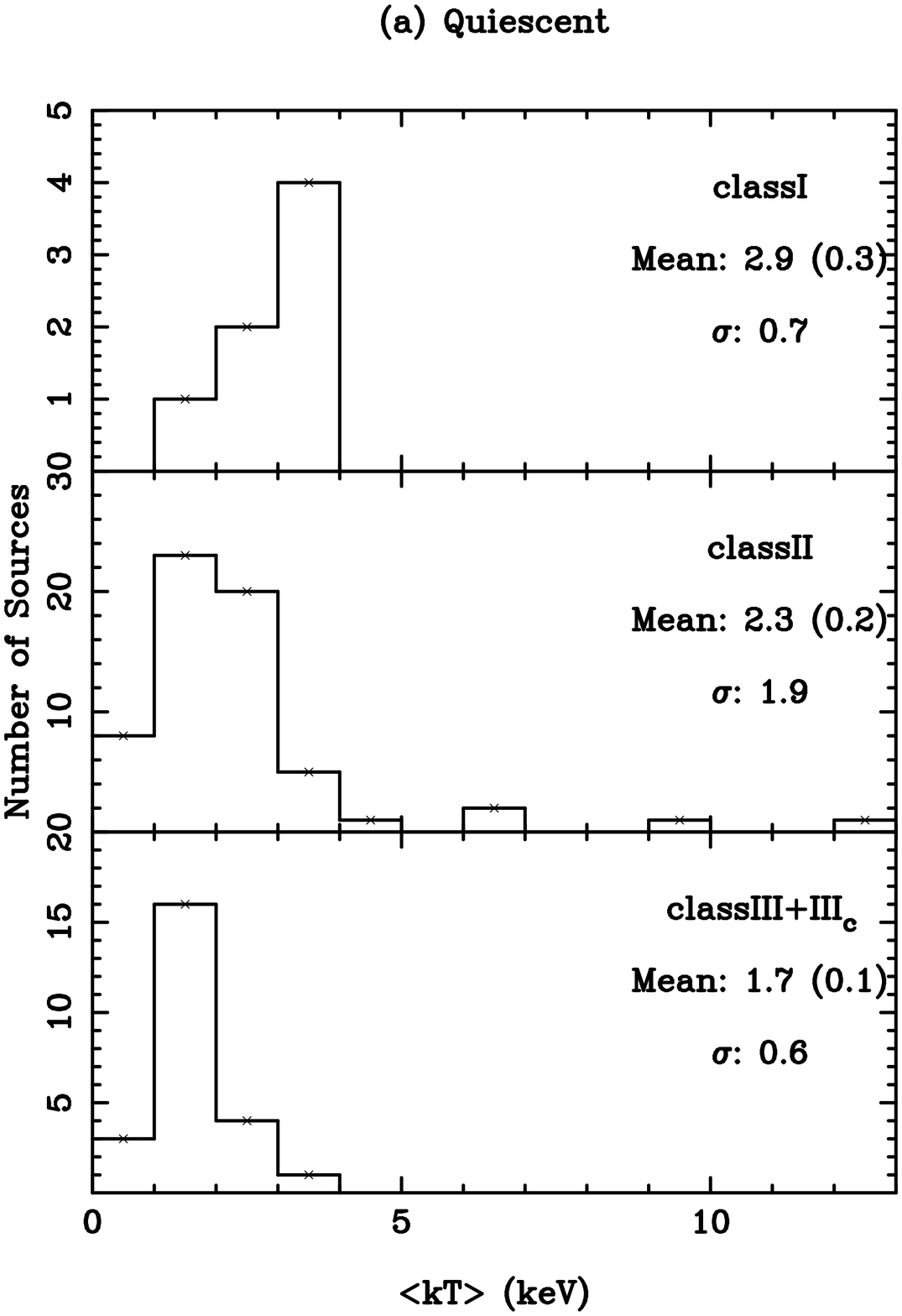}
  \hspace*{10mm}
  \FigureFile(50mm,80mm){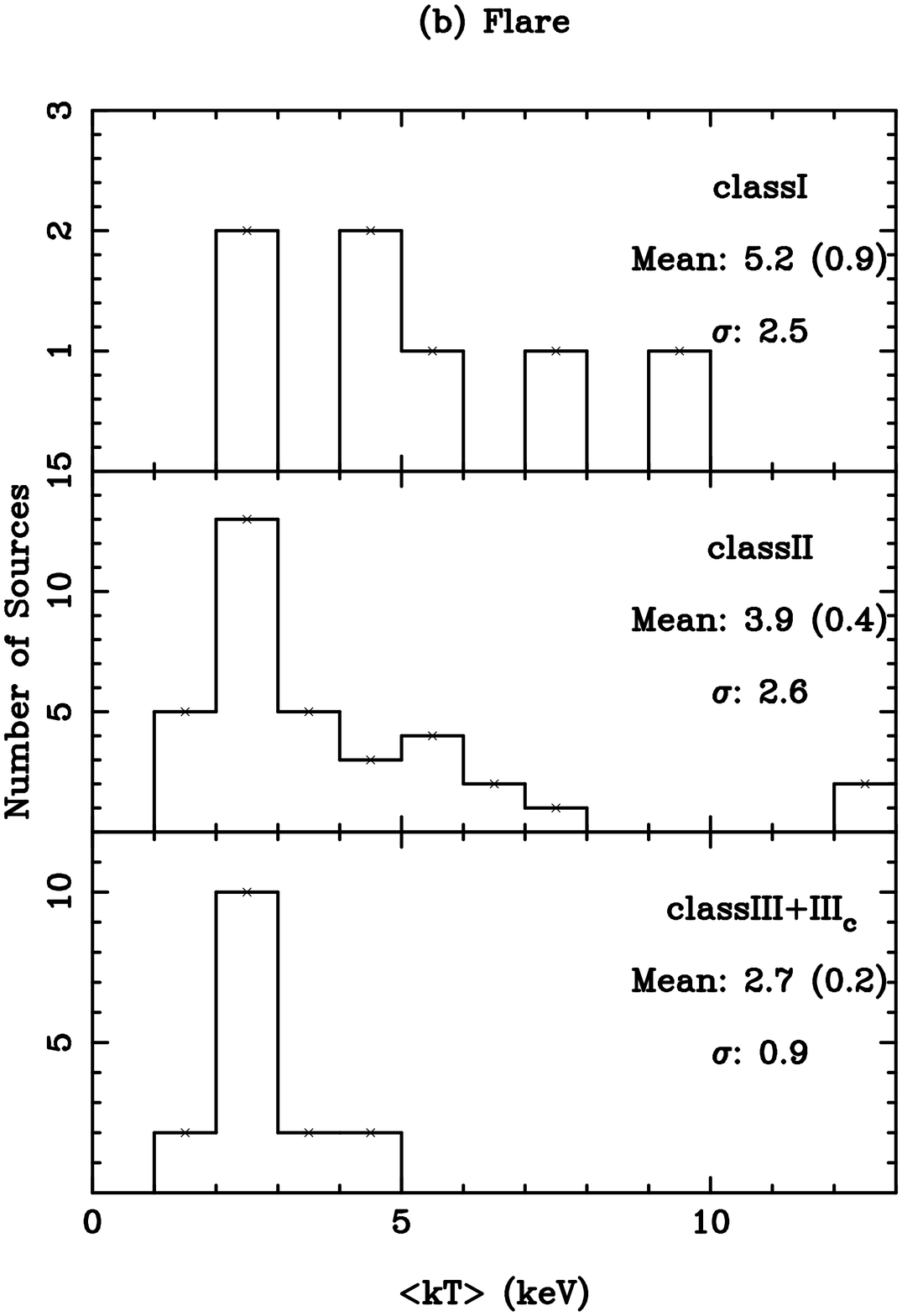}
 \end{center}
 \caption{Histograms of $kT$ in the (a) quiescent and (b) flare
 phases for each class, with their mean values (Mean) and standard
 deviation ($\sigma$) in the unit of keV for each class. The parentheses
 indicate errors of the mean values. \label{fig:hist_kt}}
\end{figure}

\begin{figure}[htbp]
 \begin{center}
  \FigureFile(50mm,80mm){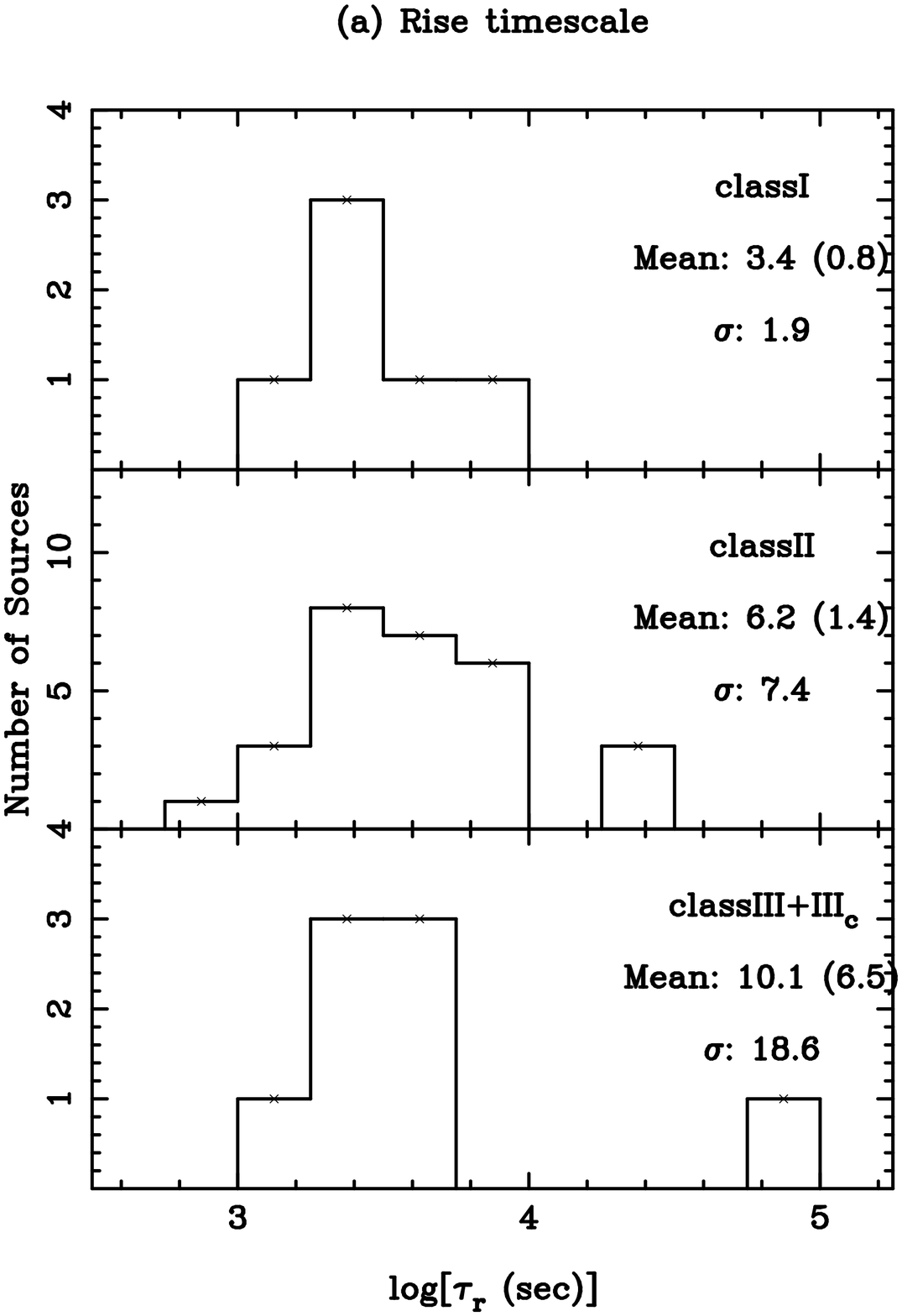}
  \hspace*{10mm}
  \FigureFile(50mm,80mm){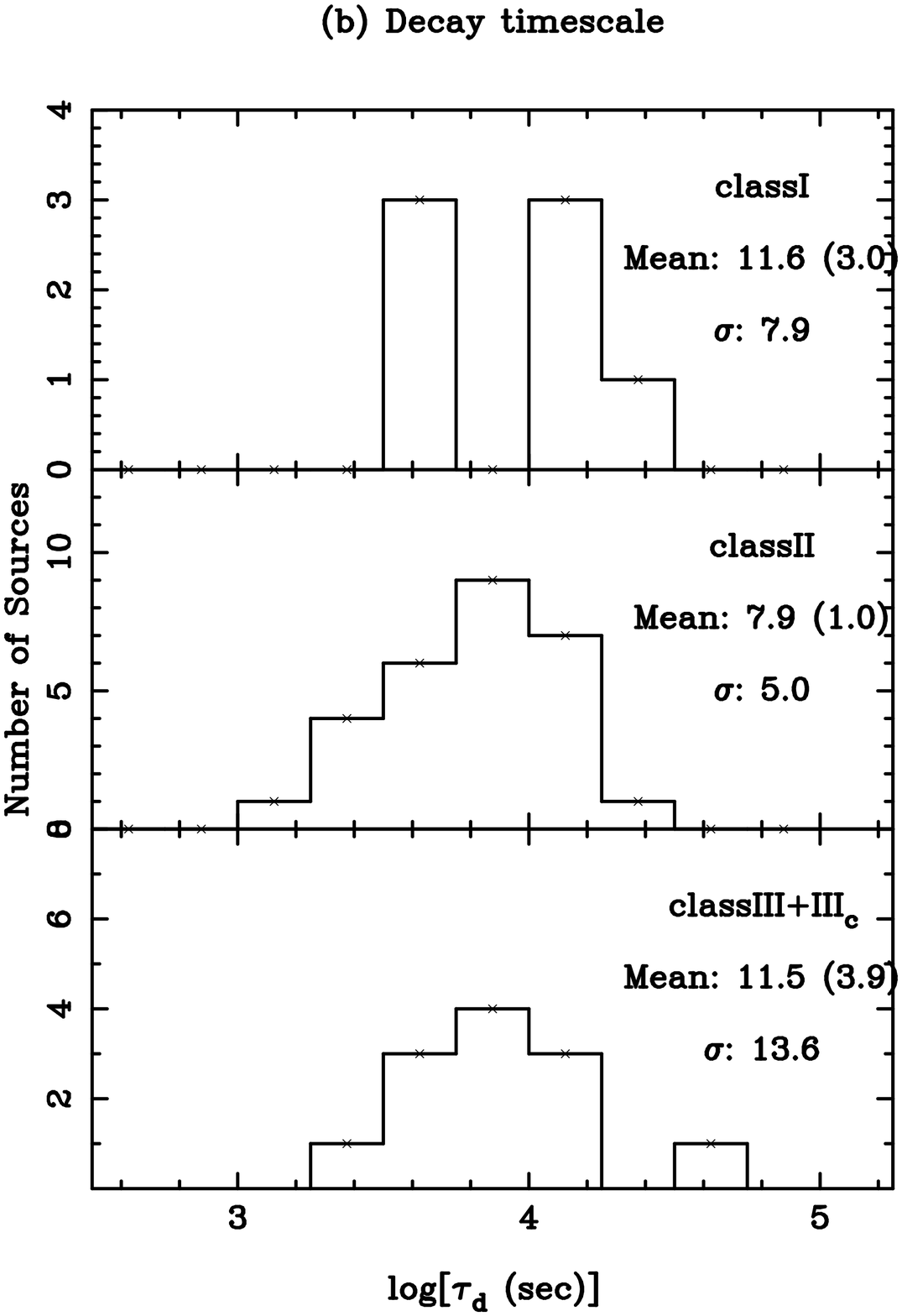}
 \end{center}
 \caption{Same as figure \ref{fig:hist_kt}, but for the flare rise and decay
 timescales, with their mean values and standard deviation in the unit
 of ks for each class. \label{fig:hist_tr_td}}
\end{figure}

In figures \ref{fig:hist_kt} and \ref{fig:hist_tr_td}, histograms of
$\langle kT\rangle$, $\tau_{\rm r}$, and $\tau_{\rm d}$ are shown for each
class separately, together with the mean values and standard deviations. 
We exclude the samples whose best-fit parameters and/or those errors were
not determined, due to the limited statistics.  For sources with
multi-temperature plasma (BF-46 and BF-96), we regard $\langle kT\rangle$
of the soft and hard components as the quiescent and flare values,
respectively, following the discussion that the former would be steady
coronal emission, while the latter is the flare activity
\citep{Imanishi2002}.

Most of the $\langle kT\rangle$ values in the quiescent phase are
in the range of
0.2--5 keV, while it becomes systematically higher (1--10 keV) in the
flare phase, indicating plasma heating during the flare.
Although the flare temperature of all classes is distributed around 3--4
keV, some flares of class I and II sources show a higher temperature than
5 keV, while all flares of class III+III$_{\rm c}$ sources have
$\langle kT\rangle$ less than 5 keV. 
%
%
The mean values of $\langle kT\rangle$ during flares for class I and II
(4--5 keV) are therefore larger than that of class III+III$_{\rm c}$ (2.7 keV). 

%
%
Like the case of $L_{\rm X}$, we estimated the significance level for the
$\langle kT\rangle$ difference among all of the classes with two nonparametric
two-sample tests in ASURV. The results are also given in table
\ref{tab:asurv}. From table \ref{tab:asurv}, we can see that both the GW and
the logrank tests for the difference of $\langle kT\rangle$
during flares between class
I+II and class III+III$_{\rm c}$ sources show high significance level of 94\%
and 98\%. One may be concerned that a systematically large absorption of class
I and II sources makes artificial spectral hardening of these
classes. We, however, argue that the observed tendency of higher
$\langle kT\rangle$ during flares for younger sources is not changed
by the extinction differences for the following reasons:\\
(1) If the intrinsic distribution of the flare $\langle kT\rangle$ in class
    III+III$_{\rm c}$ is the same as classes I and II, some of the class
    III+III$_{\rm c}$ flares should show a higher $\langle kT\rangle$
    than 5 keV, but there is no such flare in class III+III$_{\rm c}$.\\
(2) We re-estimated the mean $\langle kT\rangle$ of class I+II and class
    III+III$_{\rm c}$ during the flare for the limited sources of $N_{\rm
    H} \leq$ 5$\times$10$^{22}$ cm$^{-2}$ (most of class III+III$_{\rm c}$
    sources have lower $N_{\rm H}$ than this value). Although the
    mean $N_{\rm H}$ is nearly the same (2.5$\pm$0.2 and 2.2$\pm$0.3
    $\times$ 10$^{22}$ cm$^{-2}$ for class I+II and class
    III+III$_{\rm c}$), the mean $\langle kT\rangle$ is still higher for
    class I+II (3.7$\pm$0.4 keV) than for class III+III$_{\rm c}$
    (2.4$\pm$0.3 keV).

In the quiescent phase, on the other hand, ASURV shows a
significance level $>$95\% that the tendency of a higher $\langle kT\rangle$ of
class I than the others (table \ref{tab:asurv}). However, we suspect that
this may be a bias effect because the quiescent temperatures are
systematically low ($\lesssim 5$ keV), and hence would be significantly
affected by the absorption compared with the flare temperatures.

The rise and decay timescales of the flares ($\tau_{\rm r}$ and $\tau_{\rm d}$) are
distributed around 10$^{3.5}$ s and 10$^4$ s, respectively. Although the
mean value of $\tau_{\rm r}$ indicates that younger sources have a shorter rise
timescale, the difference of $\tau_{\rm r}$ among these classes is not
statistically significant 
%
%
($<$90\%, table \ref{tab:asurv}). The larger mean values for class
III+III$_{\rm c}$ sources are primarily due to flares with unusually long
timescales of $\gtrsim$50 ks (A-2, A-63, and BF-96);
%
%
such flares are not seen in class I and II sources ($\approx$30 ks at
the maximum).
Since we can not reject the possibility that these long timescales are
simply because of the composition of two (or more) unresolved flares, we
further estimate the mean $\tau_{\rm r}$ and $\tau_{\rm d}$ values of class
III+III$_{\rm c}$ sources without these flares to be 3.1$\pm$0.6 ks and
6.9$\pm$0.8 ks, which is comparable to those of the other classes.

\subsection{Correlation between the Flare Parameters}
\label{ssec:relation}

\begin{figure}[htbp]
 \begin{center}
  \FigureFile(80mm,50mm){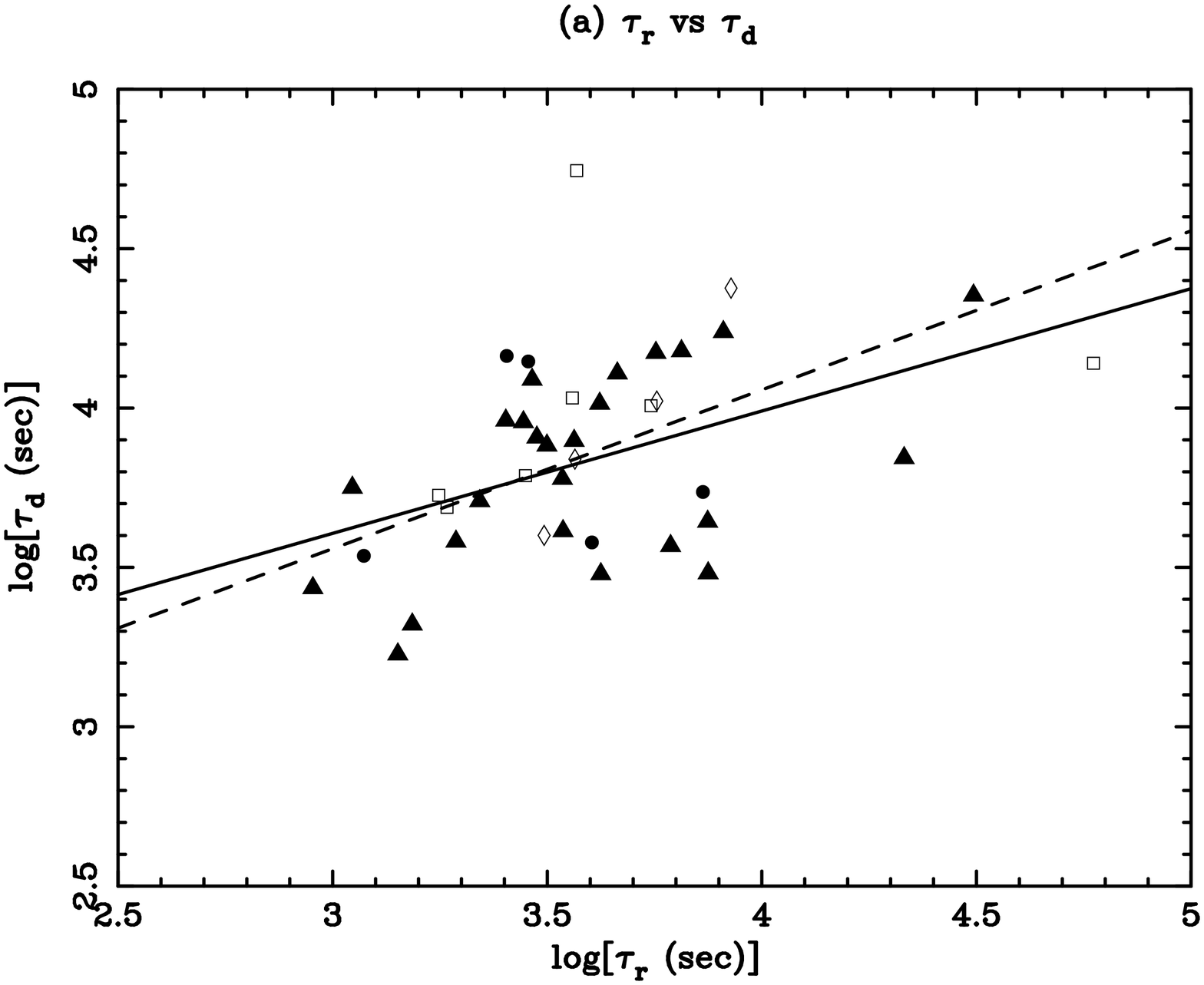}
  \FigureFile(80mm,50mm){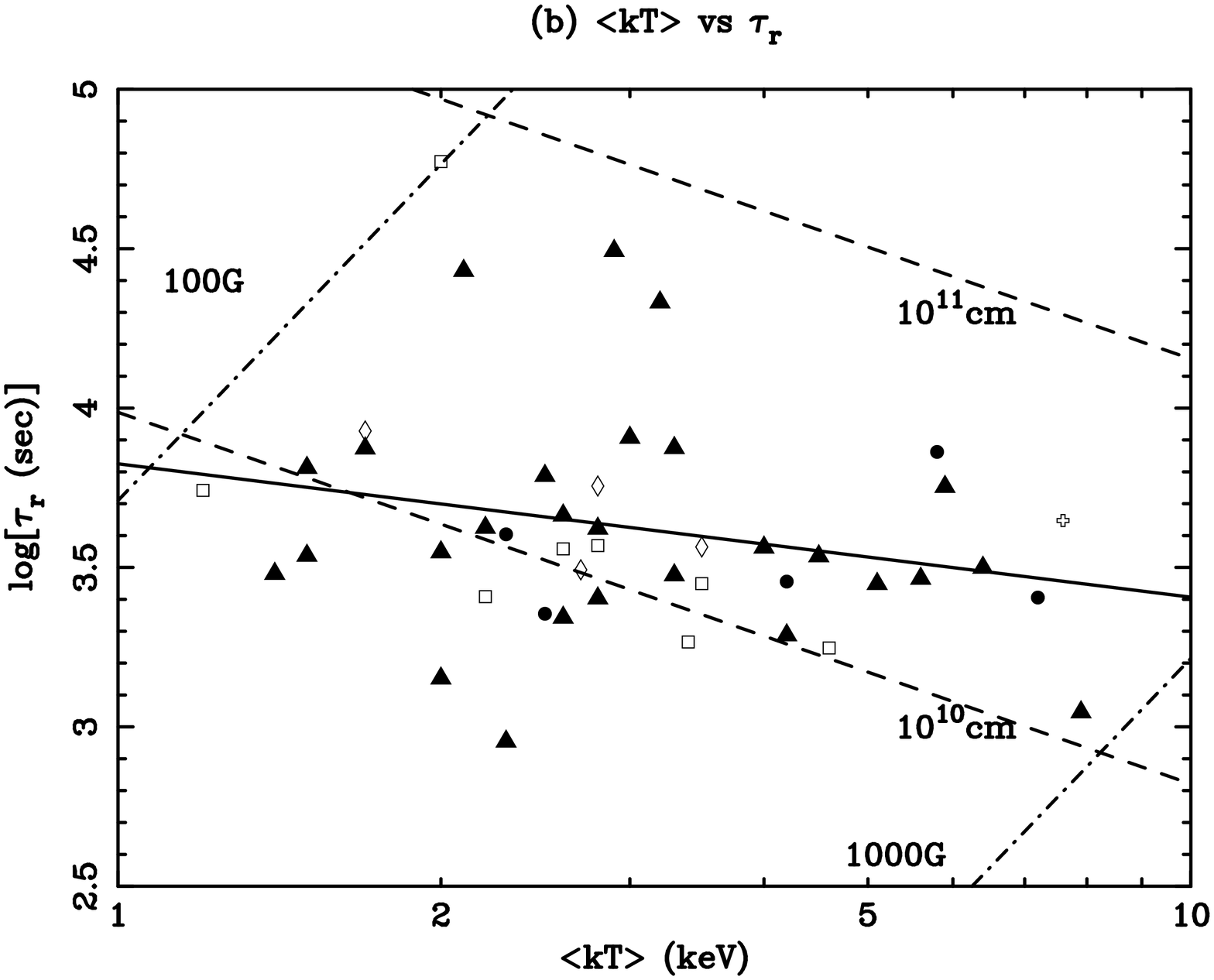}
 \end{center}
 \caption{Relations between (a) $\tau_{\rm r}$ vs $\tau_{\rm d}$ and (b) $kT$ vs
 $\tau_{\rm r}$. The circles, triangles, squares, diamonds, and crosses represent
 flares from class I, class II, class III+III$_{\rm c}$, unclassified NIR
 sources, and unidentified sources, respectively. The solid lines
 represent the best-fit log-linear correlations derived with ASURV
 [equations (\ref{eq:tr_td_fit}) and (\ref{eq:kt_tr_fit})].
 The dashed line in (a)
 is the best-fit model of equation (\ref{eq:tr_td}), while the dash-dotted and
 dashed lines in (b) are constant $B$ and $L$ lines derived by
 equations (\ref{eq:kt_tr_B}) and (\ref{eq:kt_tr_L}) with the assumption of
 $n_{{\rm c}}$ = 10$^{10.48}$ cm$^{-3}$.  \label{fig:relation}}
\end{figure}

Figure \ref{fig:relation} shows the correlations between the derived
parameters of flares; $\tau_{\rm r}$ vs $\tau_{\rm d}$ and $\langle kT\rangle$
vs $\tau_{\rm r}$. In these figures,
we exclude those flares for which we could not determine the errors
mainly due to limited statistics.
Possible positive and negative log-linear correlations can be seen in
$\tau_{\rm r}$ vs $\tau_{\rm d}$ and $\langle kT\rangle$ vs $\tau_{\rm r}$,
respectively. 
%
%
We hence checked the significances of these correlations using Cox's
proportional hazard model \citep{Isobe1986} in ASURV. This gives
significances of $\sim$99\% ($\tau_{\rm r}$ vs $\tau_{\rm d}$) and $\sim$94\%
($\langle kT\rangle$ vs $\tau_{\rm r}$). We further estimated the best-fit
log-linear models with ASURV to be
\begin{equation}
 \label{eq:tr_td_fit}
  (\frac{\tau_{\rm d}}{{\rm s}}) = 10^{2.5\pm0.4} 
  (\frac{\tau_{\rm r}}{{\rm s}})^{0.4\pm0.1} 
\end{equation}
and
\begin{equation}
 \label{eq:kt_tr_fit}
  (\frac{\tau_{\rm r}}{{\rm s}}) = 10^{3.8\pm0.1} 
  (\frac{\langle kT\rangle}{{\rm keV}})^{-0.4\pm0.3}.
\end{equation}
These models are shown by the solid lines in figure \ref{fig:relation}.

\section{Discussion}
\label{sec:discussion}

From the detected 71 X-ray flares of the $\rho$ Oph X-ray sources, we
found that (1) the flares from class I and II sources tend to show a
higher plasma temperature, which sometimes exceeds 5 keV, than class
III+III$_{\rm c}$s; (2) the distribution of $\langle L_{\rm X}\rangle$
is nearly the same among
the classes, although there is the marginal hint of slightly higher
$\langle L_{\rm X}\rangle$ for class I sources; and (3) the plots of
$\tau_{\rm r}$ vs $\tau_{\rm d}$
show positive and those of $\langle kT\rangle$ vs $\tau_{\rm r}$ show negative
log-linear correlations.
Using these enormous samples and derived properties, we discuss the
overall feature of YSO flares in this section.

\subsection{Flare Rate}
\label{ssec:flare_rate}

First, we estimate the flare rate ($F$) for each class. Following
\citet{Stelzer2000}, $F$ is defined as
\begin{equation}
 \label{eq:flare_rate}
  F = \frac{(\bar{\tau_{\rm r}} + \bar{\tau_{\rm d}}) N_{\rm F}}{N T_{\rm{obs}}} \pm 
  \frac{\sqrt{\sigma_{\tau_{\rm r}}^2 + \sigma_{\tau_{\rm d}}^2} \sqrt{N_{\rm F}}}
  {N T_{\rm{obs}}},
\end{equation}
where ($\bar{\tau_{\rm r}}$, $\bar{\tau_{\rm d}}$) and ($\sigma_{\tau_{\rm r}}$,
$\sigma_{\tau_{\rm d}}$) are the mean value and uncertainty of ($\tau_{\rm r}$,
$\tau_{\rm d}$) derived with ASURV (figure \ref{fig:hist_tr_td}), $N$ and
$N_{\rm F}$ are the number of sources and detected flares,
and $T_{\rm{obs}}$ is
the duration of the observations ($\approx$100 ks). We then determine
$F$ to be 15$\pm$1\%, 8.0$\pm$0.2\%, and 13$\pm$1\% for class I,
class II, and class III+III$_{\rm c}$ sources, respectively.
All of the values are much higher than that derived with ROSAT ($F \sim$
1\%: \cite{Stelzer2000}), which may be primarily due to the extended
sensitivity in the hard X-ray band of the Chandra observations,
because the flare activity (flux increase) is clearer in the harder
X-ray band.

The surprisingly large value of $F$ for class III+III$_{\rm c}$ sources is simply
due to the effect of flares with unusually long timescales (A-2,
A-63, and BF-96). In fact, if we regard those unusual flares as being
affected by other unknown components (unresolved flares, for example),
and exclude them, $F$ of class III+III$_{\rm c}$ becomes 5.0$\pm$0.1\%, which
is significantly smaller than those of class I--II sources. Such a
tendency is consistent with the ROSAT results ($F$ = 1.1$\pm$0.4
\% and 0.7$\pm$0.2\% for class II and III).
We hence suggest that younger sources tend to show frequent flare
activity.

\subsection{Interpretation of the $\tau_r$ vs $\tau_d$ and
$\langle kT\rangle$ vs $\tau_r$ correlations}
\label{ssec:relation_discussion}

In order to explain the observed features of the flare parameters, we
formulated $\tau_{\rm r}$, $\tau_{\rm d}$, and $\langle kT\rangle$
as a function of pre-flare
(coronal) electronic density ($n_{{\rm c}}$), half-length of the reconnected
magnetic loop ($L$) and strength of magnetic field ($B$), based on the
idea of a standard magnetic reconnection model \citep{Petschek1964} and
the balance between reconnection heating and conductive cooling
\citep{Shibata2002}. For simplicity, we assumed that $\tau_{\rm d}$ is equal to
the radiative loss timescale ($\tau_{\rm{rad}}$), although it would be
possible that $\tau_{\rm d}$ is slightly larger than $\tau_{\rm{rad}}$ (see
appendix \ref{sec:dependence}).  We also assumed that the reconnection
rate, $M_{\rm A}$ [ratio of the reconnection interval and Alfv\'en timescales;
equation (\ref{eq:tr})], is 0.01, which is the mean value of $M_{\rm A}$
for the solar observations (0.001--0.1, table 2 in \cite{Isobe2002}).
Details of the formulae are given in appendix \ref{sec:flare_model}.

\subsubsection{$\tau_{\rm r}$ vs $\tau_{\rm d}$}
\label{ssec:tr_td}

From equation (\ref{eq:tr_td}), the reconnection model predicts a positive
correlation between $\tau_{\rm r}$ and $\tau_{\rm d}$ as $\tau_{\rm d} = A
\tau_{\rm r}^{1/2}$. The slope of 1/2 shows good agreement with the observed
value of 0.4$\pm$0.1 [equation (\ref{eq:tr_td_fit})], and hence supports our
assumption that the decay phase is dominanted by radiative cooling.  From
equation (\ref{eq:tr_td}), we re-estimated log($A$) with a fixed slope of 1/2
to be 2.06$\pm$0.04 (the dashed line in figure \ref{fig:relation}a).
We also calculated log($A$) separately for each class, and then obtain
2.1$\pm$0.1, 2.0$\pm$0.1, and 2.0$\pm$0.1 for class I, II, and
III+III$_{\rm c}$, respectively. We can see no significant difference between the
classes; hence, $A$ is assumed to be the same for all sources. Since $A$
is a function of $n_{{\rm c}}$ [equation (\ref{eq:tr_td_a})], we can determine
$n_{{\rm c}}$ to be 
\begin{equation}
 \label{eq:nc}
  n_{{\rm c}} = 10^{10.48\pm0.09} (\frac{M_{\rm A}}{0.01}) \ [{\rm cm^{-3}}].
\end{equation}
Considering the possible range of $M_{\rm A}$ (0.001--0.1),
this favors a higher
pre-flare density for YSOs (10$^9$--10$^{12}$ cm$^{-3}$) than that for
the sun ($\sim$10$^9$ cm$^{-3}$).
In fact, \citet{Kastner2002} derived the plasma density of a CTTS (TW
Hydrae) using the ratio of Ne$_{\rm IX}$ and O$_{\rm VII}$ triplets, and
found an extremely high density of $\sim$10$^{13}$ cm$^{-3}$, even in their
quiescent phase. Hence, a larger $n_{{\rm c}}$ value for YSOs than that for the sun
would be a common feature.

\subsubsection{$kT$ vs $\tau_r$}
\label{ssec:kt_tr}

The predicted correlations between $\langle kT\rangle$ and $\tau_{\rm r}$
are shown in
equations (\ref{eq:kt_tr_B}) and (\ref{eq:kt_tr_L}). Positive ($\tau_{\rm r}
\propto \langle kT\rangle ^{7/2}$) and negative ($\tau_{\rm r} \propto
\langle kT\rangle ^{-7/6}$)
correlations are expected for constant $B$ and $L$ values,
respectively. The dash-dotted and dashed lines in figure
\ref{fig:relation}b show constant $B$ and $L$ lines, in which we
uniformly assume $n_{{\rm c}}$ = 10$^{10.48}$ cm$^{-3}$ [equation (\ref{eq:nc})].
Our observed negative correlation therefore predicts that flares with a higher
temperature are mainly due to a larger magnetic field. However, the
observed slope of $-$0.4$\pm$0.3 is significantly flatter than $-7/6$
[equation (\ref{eq:kt_tr_L})], and hence the effect of $L$ would not be
negligible.
We estimated the mean $B$ and $L$ for each class (table
\ref{tab:para_flare}) using the mean values of $\langle kT\rangle$ and
$\tau_{\rm r}$ (figures \ref{fig:hist_kt}b and \ref{fig:hist_tr_td}a), which are
summarized in table \ref{tab:para_flare}.
Younger sources tend to show larger $B$ values; the mean values of $B$
are $\sim$500, 300, and 200 G for class I, II, and III+III$_{\rm c}$ sources,
respectively. This is consistent with the observed result of a higher
plasma temperature for younger sources.
Also, the estimated values of $L$ (10$^{10}$--10$^{11}$ cm) suggest
that $L$ is nearly the same among these classes, comparable to the
typical radius of YSOs.
These results indicate that the flare loops for all classes are
localized at the stellar surface, which is in sharp contrast with the
idea of the flares for class I sources triggered by larger flare
loops connecting the star and the disk \citep{Montmerle2000}.

It is conceivable that the class III flares with unusually long
timescales (A-2, A-63, and BF-96) mainly affect the systematically lower
$B$ for class III sources. Since it is possible that two or more flares
make such unusually long timescales, we re-estimated $B$ and $L$ using
the mean value of $\tau_{\rm r}$ without these flares
(subsection \ref{ssec:para}). The results are shown by the parentheses in table
\ref{tab:para_flare}. Again, we confirmed the same results: class
III+III$_{\rm c}$ sources have a lower $B$ than, and comparable $L$ (but
slightly lower) to class I and II sources.

\subsection{Comparison with the $kT$--$EM$ Scaling Law}
\label{ssec:kt_em}

\begin{figure}[htbp]
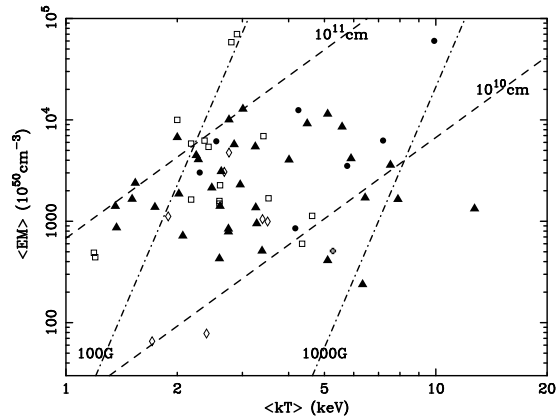

 \begin{center}
  \FigureFile(80mm,50mm){figure7.eps}
 \end{center}
 \caption{Plot of $\langle kT\rangle$ and $\langle EM\rangle$
 in the flare phases. The symbols are the
 same as in figure \ref{fig:relation}. The dashed-dotted and dashed lines
 are constant $B$ and $L$ lines derived by equations (\ref{eq:SY_B}) and
 (\ref{eq:SY_L}) with the assumption $n_{{\rm c}}$ = 10$^{10.48}$
 cm$^{-3}$. \label{fig:kt_em}}
\end{figure}

Shibata and Yokoyama (2002) showed that $L$ and $B$ of flares can be estimated
from the $kT$--$EM$ plot (see appendix \ref{sec:kt_em_appendix}). This
method is completely independent of the flare timescales ($\tau_{\rm r}$ and
$\tau_{\rm d}$), and hence can be used as a consistency check of our estimation
derived in subsection \ref{ssec:relation_discussion}.
The $\langle kT\rangle$--$\langle EM\rangle$ relation
of the $\rho$ Oph flares is shown in figure \ref{fig:kt_em}.
Using ASURV, we determined the mean values of
log($\langle EM\rangle$) to be 53.77$\pm$0.20, 53.33$\pm$0.07,
and 53.48$\pm$0.16 for class I, II, and III+III$_{\rm c}$ sources, respectively.
We then calculated the mean values of the magnetic field strength and
loop length ($B_{\rm{SY}}$ and $L_{\rm{SY}}$) by equations (\ref{eq:SY_B}) and
(\ref{eq:SY_L}) using the estimated value of $n_{{\rm c}}$ = 10$^{10.48\pm0.09}$
cm$^{-3}$ (subsubsection \ref{ssec:tr_td}). The results are given in table
\ref{tab:para_flare}.
The values of $B_{\rm{SY}}$ (100--1000 G) are much higher than those derived
in Shibata and Yokoyama (2002) and Paper I (50--150 G). This discrepancy is
caused by two different assumptions concerning the $n_{{\rm c}}$ value:
10$^9$ cm$^{-3}$ for Shibata and Yokoyama (2002) and Paper I
(typical value of the solar corona), and
10$^{10.48}$ cm$^{-3}$ for this paper.
$B_{\rm{SY}}$ and $L_{\rm{SY}}$ show good agreement with those derived in
subsubsection \ref{ssec:kt_tr} ($B$ and $L$ in table \ref{tab:para_flare}); higher
$B_{\rm{SY}}$ values for younger sources and the same order of $L_{\rm{SY}}$
comparable to the typical stellar radius.  Therefore, our estimation of
$n_{{\rm c}}$, $L$, and $B$ using the $\tau_{\rm r}$--$\tau_{\rm d}$ and
$\langle kT\rangle$--$\tau_{\rm r}$
relations is cross-checked.

The same conclusion of subsubsection \ref{ssec:kt_tr} and this section also
indicates the propriety of our assumption of $M_{\rm A}$ = 0.01. If we assume
a smaller (or larger) value of $M_{\rm A}$, $n_{{\rm c}}$ [equation (\ref{eq:nc})]
becomes smaller (larger), which causes smaller (larger) estimations of
$B_{\rm{SY}}$ [equation (\ref{eq:SY_B})], and makes a larger discrepancy
between $B$ and $B_{\rm{SY}}$ (table \ref{tab:para_flare}).

\subsection{Evolution of YSO Flares}
\label{ssec:flare_evolution}

Combining all of the results discussed in the previous sections, we propose
a simple view of the evolution of flare activity on low-mass objects as
follows.  In their earlier stage (class I), sources have a relatively
strong magnetic field ($\approx$500 G), and show frequent X-ray flares
with a higher temperature ($\approx$5 keV). As stars evolve (class II
and IIIs), the magnetic field gradually decreases (200--300 G) and makes a
moderate temperature plasma (2--4 keV) via X-ray flares. During these
phases (class I--III), the length of the flare loop does not change
significantly (10$^{10}$--10$^{11}$ cm).  As approaching to the
main-sequence stage, the magnetic field and length of the flare loop
become weak (50--150 G for the sun) and short (10$^8$--10$^9$ cm),
which causes a lower plasma temperature (0.1--1 keV) and shorter flare
timescales (10--100 s), as can be seen in the sun.

Generally, RS CVn systems show smaller flare timescales than YSOs
($\sim$100 s), while $kT$ is comparable ($\sim$5 keV). Based on the
above idea, this may be due to a larger $B$ value. In fact,
\citet{Donati1990} estimated $B$ of RS CVns to be $\approx$1000 G, which is
as large as the maximum value for YSOs (figures \ref{fig:relation}b and
\ref{fig:kt_em}).

\subsection{Comment on the Giant Flares}
\label{sec:flare_lx}

Figure \ref{fig:kt_em} indicates that the giant flares ($EM$
$\gtrsim$10$^{55}$ cm$^{-3}$) previously detected with ASCA such
as V773 Tau (class III: \cite{Tsuboi1998}) and ROXs31 (class III:
\cite{Imanishi2002}), as well as BF-64 = YLW 16A in this paper (class I;
$EM$ $\sim$10$^{55}$ cm$^{-3}$), should have either a large $L$ or $B$, if
the sustained heating is negligible.  Since the flares of ROXs31 and
YLW 16A show a relatively large rise timescale ($\sim$10$^4$ s), they may
have large $L$ values (10$^{11}$--10$^{12}$ cm). Because the rise time of V773
Tau, on the other hand, is among the shortest ($\sim$10$^{3}$ s),
the large $L_{\rm X}$ would be primarily due to the large $B$ value.  In fact,
a detailed flare decay time analysis for V773 Tau predicted an extremely
large $B$ value of $\gtrsim$1000 G \citep{Favata2001}. We suspect that
the binary nature of V773 Tau \citep{Welty1995} causes such an
exceptionally large $B$, which would be the same case as that for the
main-sequence RS CVn systems.

\section{Summary}
\label{sec:summary}

We summarize the main results of the flare analysis of two Chandra
observations of $\rho$ Oph with $\approx$100 ks exposure as follows:
\begin{enumerate}
 \item From 195 X-ray sources detected in the main region of $\rho$ Oph
       region, we found 71 X-ray flares. Most of the flares show the
       typical profile of the solar and stellar flares, while some
       bright flares have an unusually long rise timescale.
 \item Flares from class I and II sources tend to have a high plasma
       temperature, which sometimes exceeds 5 keV.
 \item There is a positive correlation between $\tau_{\rm r}$ and $\tau_{\rm d}$,
       which is well explained by the standard magnetic reconnection
       model. Shorter $\tau_{\rm r}$ and $\tau_{\rm d}$ are due to the smaller $L$
       and/or larger $B$ values.
 \item We found a negative correlation between $kT$ and $\tau_{\rm r}$, which
       indicates the same order of the flare loop length regardless of
       their classes. Larger $kT$ values for class I and II sources are
       due to larger $B$ values.
 \item The expected loop length is comparable to the stellar size,
       indicating that the scenario of the star--disk arcade magnetic
       loop is unlikely.
\end{enumerate}

\bigskip

%
%
The authors express their thanks to an anonymous referee for critical
refereeing and useful comments. The authors acknowledge Kazunari
Shibata, Takaaki Yokoyama, and Hiroaki Isobe for useful discussions and
comments. The Chandra data were obtained through the Chandra
X-ray Observatory Science Center (CXC) operated for NASA by the
Smithsonian Astrophysical Observatory. This publication makes use of
data products from the Two Micron All Sky Survey, which is a joint
project of the University of Massachusetts and the Infrared Processing
and Analysis Center/California Institute of Technology. K.I.\ and M.T.\
are financially supported by JSPS Research Fellowship for Young
Scientists.

\appendix
\section{Magnetic Reconnection Model}
\label{sec:flare_model}

In this appendix, we show details for estimating the flare
parameters ($\tau_{\rm r}$, $\tau_{\rm d}$, and $kT$), which are used in
section \ref{sec:discussion}. These formulations are based on the standard
magnetic reconnection model \citep{Petschek1964}, as well as the
observed results and analysis of solar flares.

\subsection{Estimation of the Flare Parameters}
\label{ssec:flare_model_para}

\begin{itemize}
 \item {\bf Plasma temperature ($\langle kT\rangle$)} \\
       Shibata and Yokoyama (1999; 2002) showed from MHD simulations that
       the balance between the heating rate and conduction cooling
       determines the maximum temperature of reconnection heated plasma
       ($T_{\rm{max}}$), and derived a relation for the temperature of the
       evaporated plasma ($T$) filling the reconnected magnetic loop:
       \begin{equation}
	\label{eq:kt_all0}
	 T \cong \frac{1}{3} T_{\rm{max}}
	 \cong 6.75 \times 10^7 
	 (\frac{n_{{\rm c}}}{10^9\ {\rm{cm^{-3}}}})^{-1/7} 
	 (\frac{L}{10^{11}\ {\rm{cm}}})^{2/7}
	 (\frac{B}{100\ {\rm G}})^{6/7} \ [{\rm K}],
       \end{equation}
       where $n_{{\rm c}}$, $L$, and $B$ are the pre-flare (coronal) density,
       semi-length of the loop, and magnetic field strength,
       respectively.
       As for the usual stellar flares, we could only determine the
       time-averaged temperature ($\langle kT\rangle$) due to the limited
       statistics. However, \citet{vandenOord1988} showed that the
       behavior of the flare temperature is exponential, hence we obtain
       the relation 
       \begin{equation}
	\label{eq:kt_corr}
	 T = \frac{\tau_{\rm r} + \tau_{\rm d}}
	 {\int^{0}_{-\tau_{\rm r}}e^{t/\tau_{\rm r}}dt + \int^{\tau_{\rm d}}_{0}
	 e^{-t/\tau_{\rm d}}dt} \cdot \langle T\rangle
	 \cong 1.6 \langle T\rangle .
       \end{equation}
       Using equations (\ref{eq:kt_all0}) and (\ref{eq:kt_corr}),
       $\langle kT\rangle$ is determined as
       \begin{equation}
	\label{eq:kt_all}
	 \langle kT\rangle \cong 3.64 (\frac{n_{{\rm c}}}{10^9\ {\rm{cm^{-3}}}})^{-1/7} 
	 (\frac{L}{10^{11}\ {\rm{cm}}})^{2/7}
	 (\frac{B}{100\ {\rm G}})^{6/7} \ [{\rm keV}].
       \end{equation}
 \item {\bf Rise timescale ($\tau_{\rm r}$)} \\
       Based on a standard reconnection model, $\tau_{\rm r}$ is equal
       to the interval of magnetic reconnection. \citet{Petschek1964}
       showed that the reconnection timescale is proportional to the
       Alfv\'en time, $\tau_{\rm A} \equiv L$/$v_{\rm A}$, where $v_{\rm A}$ is the
       Alfv\'en velocity. Hence $\tau_{\rm r}$ is
       \begin{equation}
	\label{eq:tr}
	 \tau_{\rm r} = \frac{\tau_{\rm A}}{M_{\rm A}} 
	 = \frac{\sqrt{4 \pi m n_{{\rm c}}}L}{M_{\rm A} B},
       \end{equation}
       where $m$ is the proton mass (=1.67$\times$10$^{-24}$ g). The
       correction factor, $M_{\rm A}$, sometimes referred to as the reconnection
       rate, was estimated to be 0.01--0.1 \citep{Petschek1964}, while
       observations for the solar flares show that $M_{\rm A}$ is in the
       range of 0.001--0.1 (table 2 in \cite{Isobe2002}), regardless of
       their flare size. $\tau_{\rm r}$ is therefore
       \begin{equation}
	\label{eq:tr_all}
	 \tau_{\rm r} \cong 1.45 \times 10^4 
	 (\frac{M_{\rm A}}{0.01})^{-1}
	 (\frac{n_{{\rm c}}}{10^{9}\ {\rm{cm^{-3}}}})^{1/2}
	 (\frac{L}{10^{11}\ {\rm{cm}}})
	 (\frac{B}{100\ {\rm G}})^{-1} \ [{\rm s}].
       \end{equation}
 \item {\bf Decay timescale ($\tau_{\rm d}$)} \\
       We assume that $\tau_{\rm d}$ is nearly the same as the radiative
       cooling timescale ($\tau_{\rm{rad}}$), i.e,
       \begin{equation}
	\label{eq:trad}
	 \tau_{\rm d} \cong \tau_{\rm{rad}} \equiv 
	 \frac{3nkT}{n^2\Lambda(T)},
       \end{equation}
       where $n$ and $\Lambda(T)$ are the maximum plasma density and
       radiative loss function given by 10$^{-24.73}$ $T^{1/4}$ for $T >
       20$ MK \citep{Mewe1985}.
       Furthermore, we assume that magnetic pressure is comparable to the
       plasma pressure at the flare peak,
       \begin{equation}
	\label{eq:Eth_EB}
	 2nkT = \frac{B^2}{8\pi}.
       \end{equation}
       Using equations (\ref{eq:kt_all0}), (\ref{eq:trad}), and
       (\ref{eq:Eth_EB}), we obtain $\tau_{\rm d}$ as
       \begin{equation}
	\label{eq:td_all}
	 \tau_{\rm d} \cong 7.75 \times 10^4
	 (\frac{n_{{\rm c}}}{10^{9}\ {\rm{cm^{-3}}}})^{-1/4}
	 (\frac{L}{10^{11}\ {\rm{cm}}})^{1/2}
	 (\frac{B}{100\ {\rm G}})^{-1/2} \ [{\rm s}].
       \end{equation}
\end{itemize}

\subsection{Predicted Correlations between the Flare Parameters}
\label{ssec:flare_model_relation}

\begin{itemize}
 \item {\bf $\tau_{\rm r}$ vs $\tau_{\rm d}$} \\
       Using equations (\ref{eq:tr_all}) and (\ref{eq:td_all}), we obtain a
       relation
       \begin{equation}
	\label{eq:tr_td}
	 (\frac{\tau_{\rm d}}{{\rm s}}) = A (\frac{\tau_{\rm r}}{{\rm s}})^{1/2},
       \end{equation}
       where
       \begin{equation}
	\label{eq:tr_td_a}
	 A \cong 640 (\frac{n_{{\rm c}}}{10^9\ {\rm{cm^{-3}}}})^{-1/2}
	 (\frac{M_A}{0.01})^{1/2}.
       \end{equation}
       Hence, a positive correlation is expected between $\tau_{\rm r}$ and
       $\tau_{\rm d}$. 
 \item {\bf $\langle kT\rangle$ vs $\tau_{\rm r}$} \\
       From equations (\ref{eq:kt_all}), (\ref{eq:tr_all}) and (\ref{eq:nc}),
       we obtain
       \begin{equation}
	\label{eq:kt_tr_B}
	 \tau_{\rm r} \cong 10^{2.20} 
	 (\frac{n_{{\rm c}}}{10^9 {\rm{cm^{-3}}}})
	 (\frac{M_{\rm A}}{0.01})^{-1}
	 (\frac{B}{100\ {\rm G}})^{-4} 
	 (\frac{\langle kT\rangle}{\rm keV})^{7/2}\ [{\rm s}],
       \end{equation}
       \begin{equation}
	\label{eq:kt_tr_L}
	 \tau_{\rm r} \cong 10^{4.82}
	 (\frac{n_{{\rm c}}}{10^9 {\rm{cm^{-3}}}})^{1/3}
	 (\frac{M_{\rm A}}{0.01})^{-1}
	 (\frac{L}{10^{11}\ {\rm{cm}}})^{4/3} 
	 (\frac{\langle kT\rangle}{\rm keV})^{-7/6}\ [{\rm s}].
       \end{equation}
       These equations indicate that the positive and negative
       log-linear correlations of $\langle kT\rangle$ vs $\tau_{\rm r}$
       are expected if $B$ and $L$ are equal for all sources, respectively.
\end{itemize}

\section{The $kT$--$EM$ Scaling Law}
\label{sec:kt_em_appendix}

Using equations (\ref{eq:kt_all0}), (\ref{eq:Eth_EB}), and assuming $V \cong
L^3$ ($V$: plasma volume), Shibata and Yokoyama (2002) derived the relation
between the flare maximum temperature ($T$) and emission measure [$EM$,
equations (5)--(6) in Shibata and Yokoyama (2002)].
Similar to equation (\ref{eq:kt_corr}), the
time behavior of $EM$ is also exponential \citep{vandenOord1988}, hence
we assume $EM \cong 1.6\langle EM\rangle$.  We thus replace equations
(5) and (6) in Shibata and Yokoyama (2002) as
\begin{equation}
 \label{eq:SY_B_corr}
  \langle EM\rangle \cong 3.7 \times 10^{48} (\frac{n_{{\rm c}}}{10^9\ {\rm{cm^{-3}}}})^{3/2} 
  (\frac{B}{100\ {\rm G}})^{-5} (\frac{\langle kT\rangle}{{\rm keV}})^{17/2}
  \ {\rm [cm^{-3}]},
\end{equation}
\begin{equation}
 \label{eq:SY_L_corr}
  \langle EM\rangle \cong 7.0 \times 10^{51} (\frac{n_{{\rm c}}}{10^9\ {\rm{cm^{-3}}}})^{2/3}
  (\frac{L}{10^{11}\ {\rm{cm}}})^{5/3} (\frac{\langle kT\rangle}{{\rm keV}})^{8/3}
  \ {\rm [cm^{-3}]}.
\end{equation}
We also derive the following two equations:
\begin{equation}
 \label{eq:SY_B}
  B \cong 52 (\frac{n_{{\rm c}}}{10^9\ {\rm{cm^{-3}}}})^{3/10} 
  (\frac{\langle EM\rangle}{10^{50}\ {\rm{cm^{-3}}}})^{-1/5} 
  (\frac{\langle kT\rangle}{{\rm keV}})^{17/10}
  \ {\rm [G]},
\end{equation}
\begin{equation}
 \label{eq:SY_L}
  L \cong 7.8 \times 10^{9} (\frac{n_{{\rm c}}}{10^9\ {\rm{cm^{-3}}}})^{-2/5}
  (\frac{\langle EM\rangle}{10^{50}\ {\rm{cm^{-3}}}})^{3/5} (\frac{\langle kT\rangle}{{\rm keV}})^{-8/5}
  \ {\rm [cm]}.
\end{equation}

\section{Possible error for the derived parameters}
\label{sec:dependence}

In the discussions of the main text (section \ref{sec:discussion}), the
assumption of $\tau_{\rm d}$ [equation (\ref{eq:trad})] includes relatively large
uncertainty. \citet{Reale1997} proposed that $\tau_{\rm d}$ becomes about a
factor of $\lesssim$ 10 larger than the radiative loss timescale
($\tau_{\rm{rad}}$), if the sustained heating exists during the decay. The
dependences of the relevant parameters on $\tau_{\rm d}$ are:
\begin{equation}
 \label{eq:depend_td_nc}
 n_{{\rm c}}    \propto \tau_{\rm d}^{-2},
\end{equation}
\begin{equation}
 \label{eq:depend_td_b}
 B      \propto n_{{\rm c}}^{1/4} \propto \tau_{\rm d}^{-1/2},
\end{equation}
\begin{equation}
 \label{eq:depend_td_l}
 L      \propto n_{{\rm c}}^{-1/2} \propto \tau_{\rm d}^{1}.
\end{equation}
Accordingly, $n_{{\rm c}}$ has large dependence on $\tau_{\rm d}$; larger $n_{{\rm c}}$ value
than $\sim$10$^{10.5}$ cm$^{-3}$ may be conceivable.
The dependence of $L$ is also relatively large. However, this
uncertainty makes $L$ much smaller, and hence larger flare loops are still
less possible. 

Another uncertainty is in the $kT$--$EM$ scaling
law. Shibata and Yokoyama (2002) showed that the effect of the filling factor
$f$ ($V$ = $f L^3$) may not be negligible. This gives possible errors
for the values in columns 5--6 of table \ref{tab:para_flare}.  The
dependences of the relevant parameters on $f$ are
\begin{equation}
 \label{eq:depend_f_lsy}
 B_{\rm{SY}} \propto EM^{-1/5} \propto f^{1/5},
\end{equation}
\begin{equation}
 \label{eq:depend_f_bsy}
 L_{\rm{SY}} \propto EM^{3/5} \propto f^{-3/5}.
\end{equation}
Hence, slightly smaller and larger values of $B_{\rm{SY}}$ and $L_{\rm{SY}}$ would
be conceivable, although the effect is only a factor of $<$ 5 (if $f$ =
0.1).


\begin{figure*}[htbp]
 \begin{center}
  \FigureFile(50mm,40mm){figure2_16.eps}
  \FigureFile(50mm,40mm){figure2_17.eps}
  \FigureFile(50mm,40mm){figure2_18.eps}
  \FigureFile(50mm,40mm){figure2_19.eps}
  \FigureFile(50mm,40mm){figure2_20.eps}
  \FigureFile(50mm,40mm){figure2_21.eps}
  \FigureFile(50mm,40mm){figure2_22.eps}
  \FigureFile(50mm,40mm){figure2_23.eps}
  \FigureFile(50mm,40mm){figure2_24.eps}
  \FigureFile(50mm,40mm){figure2_25.eps}
  \FigureFile(50mm,40mm){figure2_26.eps}
  \FigureFile(50mm,40mm){figure2_27.eps}
  \FigureFile(50mm,40mm){figure2_28.eps}
  \FigureFile(50mm,40mm){figure2_29.eps}
  \FigureFile(50mm,40mm){figure2_30.eps} \\
  Fig.\ref{fig:flares} (Continued)
 \end{center}
\end{figure*}

\begin{figure*}[htbp]
 \begin{center}
  \FigureFile(50mm,40mm){figure2_31.eps}
  \FigureFile(50mm,40mm){figure2_32.eps}
  \FigureFile(50mm,40mm){figure2_33.eps}
  \FigureFile(50mm,40mm){figure2_34.eps}
  \FigureFile(50mm,40mm){figure2_35.eps}
  \FigureFile(50mm,40mm){figure2_36.eps}
  \FigureFile(50mm,40mm){figure2_37.eps}
  \FigureFile(50mm,40mm){figure2_38.eps}
  \FigureFile(50mm,40mm){figure2_39.eps}
  \FigureFile(50mm,40mm){figure2_40.eps}
  \FigureFile(50mm,40mm){figure2_41.eps}
  \FigureFile(50mm,40mm){figure2_42.eps}
  \FigureFile(50mm,40mm){figure2_43.eps}
  \FigureFile(50mm,40mm){figure2_44.eps}
  \FigureFile(50mm,40mm){figure2_45.eps} \\
  Fig.\ref{fig:flares} (Continued)
 \end{center}
\end{figure*}

\begin{figure*}[htbp]
 \begin{center}
  \FigureFile(50mm,40mm){figure2_46.eps}
  \FigureFile(50mm,40mm){figure2_47.eps}
  \FigureFile(50mm,40mm){figure2_48.eps}
  \FigureFile(50mm,40mm){figure2_49.eps}
  \FigureFile(50mm,40mm){figure2_50.eps}
  \FigureFile(50mm,40mm){figure2_51.eps}
  \FigureFile(50mm,40mm){figure2_52.eps}
  \FigureFile(50mm,40mm){figure2_53.eps} \\
  Fig.\ref{fig:flares} (Continued)
 \end{center}
\end{figure*}

\clearpage

\begin{table}
 \begin{center}
  \caption{Log of the Chandra ACIS-I Observations on the $\rho$
  Ophiuchi Cloud \label{tab:obs}}
  \begin{tabular}{lccccc}
   \hline\hline
   Obs. ID & Sequence ID & Date & R.A.\marka &
   Decl.\marka & Exposure\markb \\
   & & & (J2000) & (J2000) & (ks) \\
   \hline
   BF & 200060 & 2000 Apr 13--14 & \timeform{16h27m18s.1} &
   \timeform{-24D34'21''.9} & 100.6 \\
   A & 200062 & 2000 May 15--17 & \timeform{16h26m35s.3} &
   \timeform{-24D23'12''.9} & \phantom{1}96.4 \\
   \hline
   %
   \multicolumn{6}{@{}l@{}}{\hbox to 0pt{\parbox{180mm}{\footnotesize
   \marka The position of the detector aimpoint (the telescope optical
   axis). \\
   \markb The live time corrected from deadtime. \\
   }\hss}}
  \end{tabular}
\end{center}
\end{table}

{\scriptsize
 \begin{longtable}
  {p{12mm}p{8mm}p{10mm}p{10mm}p{15mm}p{20mm}p{18mm}p{5mm}p{12mm}p{12mm}p{15mm}}
 \caption{Chandra X-ray Sources in the $\rho$ Oph Region.
 \label{tab:src}}
 \hline\hline
 No. & Count\marka & R.A.\markb & Decl.\markb & $\langle kT\rangle$\markc &
 log($\langle EM\rangle$)\markc & $N_{\rm H}$\markc & Flux\markd & $\langle L_{\rm X}\rangle$\markd &
 Red-$\chi^2$\marke & Comment \\
 & & (J2000) & (J2000) & (keV) & (cm$^{-3}$) & (10$^{22}$ cm$^{-2}$) & &
 & & \\
 \hline
 \endhead
 \hline
 \endfoot
 \hline
  \multicolumn{11}{l}{\hbox to 0pt{\parbox{170mm}{\footnotesize
  \marka Background-subtracted X-ray counts in 0.5--2.0 keV, 2.0--9.0
  keV, and 0.5--9.0keV for soft-band, hard-band, and the other sources,
  respectively. (m) denotes sources with marginal detections (the
  confidence level $<$ 99.9\%, see subsection \ref{ssec:src}). Although the
  confidence levels of A-48 and A-H2 are significant enough, we regard
  them as marginal sources because of the larger source size (A-48) and
  severe contamination from A-2 (A-H2). \\
  \markb Right ascension and declination for all sources are
  \timeform{16h} and \timeform{-24D}. \\
  \markc Parentheses indicate the 90\% confidence limits. \\
  \markd Observed flux (10$^{-14}$ erg s$^{-1}$ cm$^{-2}$) and
  absorption-corrected X-ray luminosity (10$^{29}$ erg s$^{-1}$) in
  0.5--9.0 keV. \\
  \marke Reduced-$\chi^2$ for the spectral fittings. Parentheses
  indicate the degrees of freedom. \\
  \markf We determine spectral parameters with fixed temperatures of 1
  keV and 5 keV (see text). For sources which only show the parameters
  for $kT$ = 1 or 5 keV, no good fitting is obtained for the other
  temperature. \\
  \markg We assume the same abundances as the ``F2'' phase in
  \citet{Imanishi2002}. The Quiescent spectrum is not obtained because
  the decay phases of the two flares occupy all of the light curve. \\
  \markh The pileup effect is not corrected. \\
  \marki No spectral fit is done due to the limited statistics. \\
  \markj Abundances are free parameters (see subsubsection 4.8.1 in Paper I and
  \cite{Imanishi_phD}). \\
  \markk We make the flare spectra with a bit larger timescale in order
  to obtain as good statistics as possible. Errors of $\langle EM\rangle$
  for A-29 in
  the quiescent are not determined because of the limited statistics. \\
  \markl The spectra show possible edge absorption of neutral Ca or warm
  Ar. The nonthermal model also well reproduces the spectra
  \citep{Hamaguchi2002}. \\
  \markm We assume the same temperature because of the limited
  statistics. \\
  \markn These show non-thermal spectra \citep{Imanishi_phD}. \\
  \marko \citet{Imanishi2002} proposed two temperature model with an
  unusual abundance pattern. \\
  \markp Foreground star. The distance is 60 pc \citep{Festin1998}. \\
  \markq The best-fit value of $\langle kT\rangle$ is not determined (larger than 10
  keV), hence we assume 10 keV temperature for the estimation of the
  other parameters. \\
  \markr $N_{\rm H}$ and reduced-$\chi^2$ are estimated by the
  simultaneous fittings with the identical sources in obs.-A.
  }}}
 \endlastfoot
A-1\markf	& 13.0(m)	&  25 55.22  	&  25 37.3  	&	1.0(fixed)      & 53.6(51.9--56.3)      & 45.5($>$6.7)          & 0.5   & 38.4  	& 0.40(2)	&  \\ 
	        &	        & 	      	& 		&	5.0(fixed)      & 51.6(50.7--53.2)      & 15.0(1.6--129.3)      & 0.6   & 0.6   	& 0.31(2)	&  \\ 
A-2\markg	& 79359.2\markh	&  26 03.02  	&  23 36.1  	&	2.9(2.8--3.0)   & 54.8(54.8--54.9)      & 1.1(1.0--1.1)         & 1257.2& 602.5 	& 1.51(436)	&  Flare 1 \\ 
	      	&		&		&		&	2.8(2.7--2.9)   & 54.8(54.8--54.8)      & ...			& 1016.0& 493.6 	& ...		&  Flare 2 \\ 
A-3	      	& 829.9    	&  26 07.05  	&  27 24.4  	&	2.1(1.8--2.7)   & 53.1(53.0--53.3)      & 3.7(3.2--4.2)         & 13.7  & 13.6  	& 0.81(76)	&  Quiescent \\ 
	        &	       	& 	      	& 		&	2.5(1.8--4.3)   & 53.3(53.1--53.5)      & ...                   & 26.1  & 22.4  	& ...		&  Flare \\ 
A-4     	& 23.7     	&  26 07.36  	&  25 31.4  	&	19.4($>$0.2)    & 51.4(51.0--59.7)      & 4.6(1.9--44.5)        & 0.8   & 0.4   	& 0.34(7)	&  \\ 
A-5     	& 994.2    	&  26 07.62  	&  27 41.6  	&	2.1(1.8--2.4)   & 53.1(53.0--53.2)      & 2.8(2.5--3.1)         & 13.4  & 11.7  	& 0.72(114)	&  Quiescent \\ 
	        &	        & 	      	& 		&	3.5(2.5--4.9)   & 53.2(53.1--53.4)      & ...                   & 34.5  & 20.8  	& ...		&  Flare 1 \\ 
	        &	        & 	      	& 		&	2.6(1.9--4.0)   & 53.2(53.0--53.3)      & ...                   & 22.0  & 16.0  	& ...		&  Flare 2 \\ 
A-6     	& 193.1    	&  26 10.34  	&  20 54.6  	&	2.3(1.4--4.3)   & 53.0(52.6--53.4)      & 5.8(4.2--8.5)         & 8.2   & 9.5   	& 0.74(17)	&  \\ 
A-7     	& 83.1     	&  26 12.42  	&  28 49.0  	&	3.3(1.4--10.3)  & 52.1(51.8--52.8)      & 5.5(3.5--10.2)        & 2.0   & 1.6   	& 0.57(24)	&  \\ 
A-8     	& 16.2     	&  26 13.30  	&  28 23.0  	&	2.0($>$0.2)     & 51.4(50.6--56.1)      & 3.0(0.8--10.2)        & 0.3   & 0.3   	& 0.95(2)	&  \\ 
A-9\markf     	& 12.9     	&  26 14.34  	&  22 28.1  	&	1.0(fixed)      & 53.0(51.4--54.1)      & 26.6(3.4--92.8)       & 0.2   & 8.9   	& 0.57(2)	&  \\ 
	        &	        & 	      	& 		&	5.0(fixed)      & 51.4(50.4--51.9)      & 10.8(0.8--46.5)       & 0.4   & 0.3   	& 0.39(2)	&  \\ 
A-10\marki    	& 10.9      	&  26 15.06  	&  25 48.1  	&	...             & ...                   & ...                   & ...   & ...   	& ...		&  \\ 
A-11    	& 59.8     	&  26 15.71  	&  27 49.3  	&	3.7($>$0.7)     & 52.0(51.6--54.2)      & 6.6(3.3--25.6)        & 1.5   & 1.3   	& 0.56(14)	&  \\ 
A-12    	& 180.3    	&  26 15.81  	&  19 22.2  	&	1.2(1.1--1.6)   & 52.6(52.3--52.9)      & 2.6(2.0--3.0)         & 1.8   & 3.1   	& 0.73(15)	&  Quiescent \\ 
	        &	        & 	      	& 		&	4.3($>$0.9)     & 52.8(52.4--53.3)      & ...                   & 15.1  & 8.1   	& ...		&  Flare \\ 
A-13    	& 16.2     	&  26 16.32  	&  18 43.0  	&	5.2($>$0.5)     & 51.4(50.9--55.4)      & 5.8(1.6--43.6)        & 0.5   & 0.3   	& 0.01(1)	&  \\ 
A-14    	& 2068.5   	&  26 16.87  	&  22 23.0  	&	1.7(1.5--1.9)   & 53.3(53.2--53.3)      & 1.9(1.8--2.1)         & 18.6  & 16.3  	& 1.14(165)	&  Quiescent \\ 
	        &	        & 	      	& 		&	2.6(2.2--3.1)   & 53.5(53.4--53.6)      & ...                   & 54.0  & 33.4  	& ...		&  Flare \\ 
A-15    	& 3741.7   	&  26 17.06  	&  20 21.6  	&	1.3(1.2--1.3)   & 53.1(53.1--53.1)      & 0.5(0.5--0.6)         & 19.0  & 10.7  	& 1.29(181)	&  Quiescent \\ 
	        &	        & 	      	&		&	1.4(1.3--1.5)   & 53.1(53.1--53.2)      & ...                   & 22.5  & 12.1  	& ...		&  Flare 1 \\ 
	        &	        & 	      	&		&	1.5(1.3--1.7)   & 53.4(53.3--53.4)      & ...                   & 41.4  & 20.7  	& ...		&  Flare 2 \\ 
	        &	        & 	      	&		&	1.5(1.4--1.6)   & 53.2(53.2--53.3)      & ...                   & 28.4  & 14.3  	& ...		&  Flare 3 \\ 
A-16    	& 126.2    	&  26 17.13  	&  12 38.9  	&	0.9(0.6--1.2)   & 52.4(52.2--52.8)      & 1.8(1.4--2.2)         & 1.1   & 2.5   	& 1.26(16)	&  \\ 
\end{longtable}
}

{\footnotesize
\begin{longtable}
  {p{10mm}p{10mm}p{20mm}p{8mm}p{5mm}p{5mm}p{5mm}p{50mm}p{15mm}}
 \caption{Identifications of the X-ray sources. \label{tab:id}}
 \hline\hline
 No. & Offset\marka & Radio\markb & \multicolumn{4}{c}{---- X-ray
----\markc} & Other names\markd & Class\marke \\
  & ($''$) & & PSPC & HRI & ASCA & IKT & & \\
 \hline
 \endhead
 \hline
 \endfoot
 \hline
 \multicolumn{9}{l}{\hbox to 0pt{\parbox{170mm}{\footnotesize
 \marka Offset between the Chandra and nearest 2MASS sources. For
 sources having a radio counterpart only, the offset between the {\it
 Chandra} and radio sources \citep{Andre1987, Leous1991} are shown,
 which is indicated by (R). \\
 \markb Radio sources. S, R, and L indicate sources listed in
 \citet{Andre1987} (ROC), \citet{Stine1988} (SFAM), and
 \citet{Leous1991} (LFAM), respectively. Lp denotes possible radio
 sources in LFAM. \\
 \markc X-ray sources detected with ROSAT/PSPC
 \citep{Casanova1995}, ROSAT/HRI \citep{Grosso2000}, and
 ASCA \citep{Kamata1997}. ``IKT'' denotes Chandra sources already
 listed in Paper I. \\
 %
%
%
 \markd Source names used in the literature. Abbreviations for names are
 SR \citep{Struve1949}, DoAr \citep{Dolidze1959}, ROXs
 \citep{Bouvier1992}, S \citep{Grasdalen1973}, YLW \citep{Young1986}, GY
 \citep{Greene1992}, WL \citep{Wilking1983}, VSSG \citep{Vrba1975}, GSS
 \citep{Grasdalen1973}, Elias \citep{Elias1978}, BBRCG
 \citep{Barsony1989}, CRBR \citep{Comeron1993}, SKS \citep{Strom1995},
 IRS \citep{Wilking1989}, and ISO \citep{Bontemps2001}. For sources
 having NIR counterpart only in the 2MASS catalog, we alternatively show
 the 2MASS source names. \\
 \marke Source classifications; I:~class I, II:~class II, III:~class
 III, III$_{\rm c}$:~class III candidate (table 5 in \cite{Bontemps2001}),
 BD:~brown dwarf, BD$_{\rm c}$:~brown dwarf candidate \citep{Imanishi2001b},
 F:~foreground star \citep{Festin1998}. Unclassified NIR sources are
 indicated by ``?'', and sources with no NIR counterpart are called
 ``unidentified sources'', labeled no data point (...) in this
 column. For BD/BD$_{\rm c}$s, we show the available IR classification in
 \citet{Bontemps2001} in the parentheses. \\
 \markf Offset from a nearest NIR source in Greene and Young (1992). \\
 \markg \citet{Grosso2001}. \\
 \markh \citet{Tsuboi2000}. \\
 \marki The position of GY5 \citep{Greene1992} is slightly
 shifted ($\sim$2\farcs3) from the 2MASS source. \\
 \markj A candidate of HH object \citep{Gomez1998}. \\
 %
%
%
 \markk From NIR spectroscopy, \citet{Wilking1999} derived for these M
 dwarfs masses higher than the hydrogen burning limit.
 }}}
 \endlastfoot
%
%
A-1	& ...		& ...		& ...		& ...		& ...		& ...	&	...									& ...			\\	
A-2	& 0.33		& R8, S7 	& 13		& A11		& 3		& ...	&	DoAr21, ROXs8, YLW26, GSS23, Elias14, SKS1-5, ISO10			& III           	\\	
A-3	& 0.40		& ...		& ...		& ...		& ...		& ...	&	ISO13									& II            	\\	
A-4	& 7.54(R)	& R9?		& ...		& ...		& ...		& ...	&	...									& ...			\\	
A-5	& 0.60		& ...		& 15?		& A13		& ...		& ...	&	ISO14									& III           	\\	
A-6	& 0.23		& ...		& 16		& ...		& ...		& ...	&	GSS26, CRBR5, SKS1-6, ISO17						& II			\\	
A-7	& ...		& ...		& ...		& ...		& ...		& ...	&	...									& ...			\\	
A-8	& ...		& ...		& ...		& ...		& ...		& ...	&	...									& ...			\\	
A-9	& ...		& ...		& ...		& ...		& ...		& ...	&	...									& ...			\\	
A-10   	& ...		& ...		& ...		& ...		& ...		& ...	&	...									& ...			\\	
A-11	& ...		& ...		& ...		& ...		& ...		& ...	&	...									& ...			\\	
A-12	& 0.47		& ...		& ...		& ...		& ...		& ...	&	CRBR9, SKS1-7, ISO18							& III$_{\rm c}$       	\\	
A-13	& ...		& ...		& ...		& ...		& ...		& ...	&	...									& ...			\\	
A-14	& 0.11		& Lp1		& 18		& A14		& ...		& ...	&	ROXs11, GSS29, Elias18, CRBR10, SKS1-8, ISO19                  		& II            	\\	
A-15	& 0.44		& ...		& 17		& A15		& ...		& ...	&	DoAr24, ROXs10A, GSS28, Elias19, CRBR11, SKS1-9, ISO20         		& II            	\\	
A-16	& 1.08		& ...		& ...		& ...		& ...		& ...	&	2MASSI J1626172$-$241238						& ?	            	\\	
\hline
\end{longtable}
}

\begin{longtable}[htbp]{lccccc}
 \caption{Detected X-ray Flares. \label{tab:flare}}
 \hline\hline
 No. & $t_{\rm p}$ & $\tau_{\rm r}$ & $\tau_{\rm d}$ & $Q$ & Comment \\ 
 & (ks) & (ks) & (ks) & (count ks$^{-1}$) \\
 \hline
 \endhead
 \hline
 \endfoot
 \hline
 \multicolumn{6}{l}{\hbox to 0pt{\parbox{140mm}{\footnotesize
 \marka We can not determine errors because of the limited statistics
 and/or strong coupling of the other parameters. \\
 \markb Fitting is obtained for light curves in 2\farcs5--12\farcs5
 (A-2) and 2\farcs5--7\farcs5 (BF-64) radius circles in order to avoid
 the pileup effect. \\
 \markc No fitting is possible because of the limited statistics and/or
 the boundary of the observations. \\
 \markd We relax the flare peak time ($t_{\rm p}$) to be free in order to
 obtain as good results as possible. \\
 \marke This should be the upper limit because the fitting does not
 include the time bin with zero counts. 
 }}}
 \endlastfoot
 \multicolumn{6}{c}{--- Class I ---} \\
A-49		& 58	& 0.9\marka		& 12.2(9.2--16.0)	& 0.8(0.5--1.0)\marke	& \\ 
BF-26		& 18	& 7.3(6.0--9.0)		& 5.5(4.4--6.7)		& 7.0(6.4--7.5)		& \\ 
BF-61		& 82	& 2.9(2.6--3.2)		& 14.0(12.7--15.6)	& 10.7(10.0--11.3)	& \\ 
BF-64		& 22	& 2.5(2.0--3.2)		& 14.6(13.2--16.1)	& 11.4(9.7--13.0)	& Flare 1 \\ 
\markb		& 87.3(87.1--87.6)\markd	
			& 5.1\marka		& 27.7(24.3--31.5)	& ...			& Flare 2 \\ 
BF-89		& 4	& 1.2(0.8--1.6)		& 3.4(2.9--4.1)		& 6.2(5.6--6.7)		& Flare 1 \\ 
		& 88	& 4.0(2.4--6.2)		& 3.8(2.1--7.5)		& ...			& Flare 2 \\ 
		& ...\markc	
			& 2.3(1.5--3.4)		& ...\markc		& ...			& Flare 3 \\ 
\multicolumn{6}{c}{--- Class II ---} \\
A-3		& 24	& 6.1(4.3--8.6)		& 3.7(1.8--9.3)		& 7.0(6.4--7.5)		& \\ 
A-14		& 8	& 4.6(3.3--6.7)		& 12.8(10.6--15.7)	& 14.6(13.6--15.6)	& \\ 
A-15		& 2	& ...\markc		& 11.7(4.6--28.2)	& 28.1(22.4--30.8)	& Flare 1 \\ 
		& 60	& 3.4(2.3--5.2)		& 4.1(3.0--5.8)		& ...			& Flare 2 \\ 
		& 86	& 6.5(3.7--12.1)	& 15.1(8.1--48.0)	& ...			& Flare 3 \\ 
A-22		& 20	& 4.2(2.8--6.1)		& 10.3(6.3--16.5)	& 7.5(6.7--8.2)		& \\ 
A-23		& 56	& 0.8\marka		& 4.0(3.3--4.8)		& 2.0(1.7--2.3)		& Flare 1 \\ 
		& 74	& 1.9(0.8--3.4)		& 3.8(2.1--6.3)		& ...			& Flare 2 \\ 
A-24		& 72	& 2.9(2.7--3.1)		& 12.3(11.3--13.3)	& 6.4(5.9--7.0)		& \\ 
A-25		& 10	& 0.9\marka		& 1.3\marka		& 17.0(15.9--18.0)	& Flare 1 \\ 
		& 38	& 7.5(4.2--12.3)	& 3.0(0.7--10.1)	& ...			& Flare 2 \\ 
		& 88	& 1.1(0.5--1.7)		& 5.6(3.8--9.1)		& ...			& Flare 3 \\ 
A-34		& 36	& 3.0(1.5--5.8)		& 8.1(5.6--11.8)	& 2.4(2.0--2.8)		& \\ 
A-53		& 82	& 2.2(1.2--3.6)		& 5.1(2.8--10.0)	& 10.6(10.0--11.3)	& \\ 
A-69		& 70	& 26.9(11.0--121.3)	& 4.7\marka		& 3.0\marka		& \\ 
A-77		& 98	& 3.5(2.2--5.5)		& ...\markc		& 0.7(0.5--0.8)\marke	& \\ 
A-78		& 94	& 1.5\marka		& 3.1\marka		& 1.0(0.8--1.1)		& \\ 
A-79		& 42	& 2.8(2.5--3.0)		& 9.0(20.8--16.3)	& 2.4(1.7--3.1)		& Flare 1 \\ 
		& 62	& 8.1(6.9--9.2)		& 17.3(35.9--33.4)	& ...			& Flare 2 \\ 
BF-8		& 66	& 1.0\marka		& 1.3\marka		& 0.6(0.3--0.8)\marke	& \\ 
BF-17		& 32	& 1.4(0.8--2.1)		& 1.7(0.8--2.9)		& 1.0(0.8--1.1)		& \\ 
BF-27		& 86	& 1.5(0.6--2.5)		& 2.1(1.1--3.4)		& 0.9(0.7--1.1)		& \\ 
BF-28		& 18	& 0.9(0.2--1.3)		& 2.7(1.9--4.0)		& 1.1(0.9--1.3)		& \\ 
BF-31		& 54	& 2.5(2.3--2.7)		& 9.1(8.4--10.0)	& 1.1(0.8--1.3)		& \\ 
BF-35		& 61	& 6.0(5.2--6.9)		& 15.8(13.7--18.2)	& 2.8(2.3--3.2)		& \\ 
BF-42		& 76	& 7.5(3.8--15.1)	& 4.4(1.7--7.6)		& 3.8(3.4--4.2)		& \\ 
BF-51		& 5	& 1.0\marka		& 1.0\marka		& 0.6(0.4--0.8)\marke	& Flare 1 \\ 
		& 61	& 1.5\marka		& 1.3\marka		& ...			& Flare 2 \\ 
BF-59		& 88	& 3.2(2.7--3.6)		& 7.6(6.2--9.5)		& 2.0(1.7--2.3)		& \\ 
BF-63		& 54	& ...\markc		& 6.9(5.3--8.9)		& 2.9(2.6--3.2)		& \\ 
BF-66		& 98	& 3.0(1.2--5.8)		& 0.5\marka		& 2.4(2.1--2.7)		& \\ 
BF-78		& 4	& 3.7(2.5--5.6)		& 7.9(6.8--9.2)		& 21.8(20.9--22.8)	& \\ 
BF-87		& 84	& 3.4(3.1--3.8)		& 6.0(5.4--6.7)		& 1.0(0.8--1.2)		& \\ 
BF-88		& 14	& 21.5(10.7--56.0)	& 7.0(3.6--15.6)	& 26.0(14.7--32.7)	& Flare 1 \\ 
		& 46	& 31.1(17.9--52.5)	& 22.6(14.7--32.7)	& ...			& Flare 2 \\ 
		& 84	& 4.2(1.9--20.2)	& 3.0(1.3--9.1)		& ...			& Flare 3 \\ 
\multicolumn{6}{c}{--- Class III ---} \\
A-2\markb	& ...\markc	& ...\markc	& 131\marka		& ...\markc		& Flare 1 \\ 
		& 70	& 3.7(2.1-5.8)		& 55.6(30.8--117.5)	& ...			& Flare 2 \\ 
A-5		& 54	& 2.8(1.8--4.1)		& 6.1(4.0--9.5)		& 8.0(7.2--8.7)		& Flare 1 \\ 
		& 76	& 2.3\marka		& 7.0(2.6--22.9)	& ...			& Flare 2 \\ 
A-20		& 22	& 17.0\marka		& 8.3(3.4--31.3)	& 27.3\marka		& Flare 1 \\ 
		& 62	& 1.4\marka		& 2.7(2.7--2.7)		& ...			& Flare 2 \\ 
		& 76	& 4.2\marka		& 9.9(5.6--20.8)	& ...			& Flare 3 \\ 
A-41		& 90	& 2.0\marka		& ...\markc		& 21.6(20.7--22.4)	& \\ 
A-63		& 52	& 2.6(1.3--4.3)		& 66.4\marka		& 10.3(9.5--11.1)	& \\
A-81		& 17	& 1.8(1.4--2.2)		& 5.3(4.2--6.7)		& 1.4(1.2--1.7)		& \\ 
BF-40		& 88	& 1.8(1.2--2.7)		& 4.9(3.8--6.4)		& 9.7(9.1--10.2)	& \\ 
BF-46		& 72	& 3.6(2.8--4.7)		& 10.8(8.1--14.2)	& 76.5(74.6--78.3)	& \\ 
BF-84		& 72	& 5.5(2.7--12.3)	& 10.2(3.3--24.9)	& 1.0(0.7--1.2)		& \\ 
BF-96		& 32	& 59.2(41.9--96.7)	& 13.8(10.8--17.9)	& 27.2(24.8--29.2)	& \\ 
\multicolumn{6}{c}{--- Class III$_c$ ---} \\
A-12		& 20	& 0.9\marka		& 0.9\marka		& 1.3(1.1--1.5)		& \\ 
BF-72		& ...\markc	
			& ...\markc		& 4.1(2.1--6.8)		& 1.2(1.0--1.4)		& \\ 
\multicolumn{6}{c}{--- Unclassified NIR sources ---} \\
A-26		& 2	& 3.1(1.0--57.9)	& 4.0(3.1--5.1)		& 1.4(1.0--1.7)		& Flare 1 \\ 
		& 18	& 1.3\marka		& 4.6(2.3--10.9)	& ...			& Flare 2 \\ 
		& 76	& ...\markc		& 20.1\marka		& ...			& Flare 3 \\ 
A-28		& 50	& 8.5(4.3--17.8)	& 23.8(9.1--55.4)	& 0.8(0.3--1.1)		& \\ 
A-76		& 72	& 1.9\marka		& 24.0(12.2--129.1)	& 1.0(0.8--1.2)		& \\ 
BF-55		& 30 	& 1.5\marka		& 0.8\marka		& 0.7(0.5--0.9)\marke	& \\ 
BF-62		& 20	& 3.7(2.6--5.0)		& 6.9(4.9--9.9)		& 1.2(0.5--1.8)		& Flare 1 \\ 
		& 53	& 5.7(5.0--6.5)		& 10.5(9.3--12.0)	& ...			& Flare 2 \\ 
\multicolumn{6}{c}{--- Unidentified sources ---} \\
A-29		& 80	& 1.0\marka		& 2.1\marka		& 0.5\marka\marke	& \\ 
BF-36		& 30	& 4.4(1.7--9.4)		& ...\markc		& 0.8(0.6--0.9)		& \\ 
BF-92		& 20	& 0.5\marka		& 2.6(1.1--4.4)		& 1.9\marka		& \\ 
\hline
\end{longtable}

%
%
\begin{table}
 \begin{center}
  \caption{Results of two-sample tests for the differences of the
  observed parameters among classes. \label{tab:asurv}}
  \begin{tabular}{lccccccccc}
   \hline\hline
   Parameter & \multicolumn{3}{c}{--- Sample size\marka ---} & 
   \multicolumn{6}{c}{--- Probability for the null hypothesis\markb ---}
   \\
   & I & II & III+III$_{\rm c}$ & \multicolumn{2}{c}{I vs II+III+III$_{\rm c}$} &
   \multicolumn{2}{c}{I+II vs III+III$_{\rm c}$} & \multicolumn{2}{c}{II vs
   III+III$_{\rm c}$} \\
   & & & & GW & logrank & GW & logrank & GW & logrank \\
   \hline
   log$L_{\rm X}$ (Flare) & 8 & 36 & 16 & 0.059 & 0.063 & ... & ... &
   ... & ... \\
   log$L_{\rm X}$ (Quiescent) & 8 & 61 & 26 & ... & ... & ... &
   ... & ... & ... \\
   $\langle kT\rangle$ (Flare) & 7 & 35 & 16 & 0.079 & ... & 0.060 & 0.017 &
   ... & 0.047 \\
   $\langle kT\rangle$ (Quiescent) & 7 & 61 & 24 & 0.008 & 0.022 & 0.046 & 0.037 &
   ... & 0.026 \\
   $\tau_{\rm r}$ & 6 & 28 & 8 & ... & ... & ... & ... &
   ... & ... \\
   $\tau_{\rm d}$ & 7 & 28 & 12 & ... & ... & ... & ... &
   ... & ... \\
   \hline
   \multicolumn{10}{@{}l@{}}{\hbox to 0pt{\parbox{150mm}{\footnotesize
   \marka The number of data points used for the two-sample tests. \\
   \markb Probability that the hypothesis of two distributions
   being the same is true, which is derived with the Gehan's generalized 
   Wilcoxon test (GW) and the logrank test \citep{Feigelson1985}. Blank
   (...) indicates that the probability is larger than 0.1. \\
   }\hss}}
  \end{tabular}
 \end{center}
\end{table}

\begin{table}
 \begin{center}
  \caption{Estimated Mean Values of the Flare Physical Parameters.
  \label{tab:para_flare}}
  \begin{tabular}{lccccc}
   \hline\hline
   Class & \multicolumn{3}{c}{--- This work ---} & 
   \multicolumn{2}{c}{--- $kT$ vs $EM$ ---} \\
   & log($n_{{\rm c}}$)\markb\markc & log($B$)\markb & log($L$)\markb &
   log($B_{\rm{SY}}$)\markd & log($L_{\rm{SY}}$)\markd \\
   & (cm$^{-3}$) & (G) & (cm) & (G) & (cm) \\
   \hline
I		& 10.48$\pm$0.09& 2.7$\pm$0.1		& 10.3$\pm$0.1	& 2.6$\pm$0.1	& 10.4$\pm$0.2  \\
II		& ...		& 2.5$\pm$0.1		& 10.4$\pm$0.1	& 2.5$\pm$0.1	& 10.4$\pm$0.1  \\
III+III$_c$	& ...		& 2.3$\pm$0.1		& 10.4$\pm$0.2	& 2.2$\pm$0.1	& 10.7$\pm$0.1  \\
		&		& (2.4$\pm$0.1)\marke	& (10.0$\pm$0.1)\marke	&	&		  \\
   \\
   Equation\marka & (\ref{eq:nc}) &(\ref{eq:kt_tr_B}) &
   (\ref{eq:kt_tr_L}) & (\ref{eq:SY_B}) & (\ref{eq:SY_L}) \\
   \hline
   \multicolumn{6}{@{}l@{}}{\hbox to 0pt{\parbox{120mm}{\footnotesize
   \marka Equations in the text used for the estimation of each
   parameter. \\
   \markb The values are derived by assuming $M_{\rm A}$ = 0.01 \\
   \markc We assume the same values of $n_{{\rm c}}$ for all classes (see
   text). \\
   \markd These values are estimated using the derived values of $n_{{\rm c}}$
   (= 10$^{10.48}$ cm$^{-3}$, column 2). \\
   \marke The mean values for class III+III$_{\rm c}$ when we exclude the
   flares with unusually long timescales (A-2, A-63, and BF-96). \\
   }\hss}}
  \end{tabular}
 \end{center}
\end{table}

\end{document}